# Systematic modulation of superconducting gap dynamics in YBCO|BNT|YBCO Josephson Junctions through THz field interaction and BNT ferroelectric barrier

Sukhanidhan Singh[c*], Muthukkumaran Karthikeyan[ab], Geoffrey Chanda[de], Thanayut Kaewmaraya[fg], Pairot Moontragoon[fg], Guoxing Sun[h], Zongjin Li[i], Anucha Watcharapasorn[abc*]

[a]Center of Excellence in Quantum Technology, Faculty of Engineering, Chiang Mai University, Chiang Mai, 50200, Thailand

[b]Department of Physics and Materials Science, Faculty of Science, Chiang Mai University, Chiang Mai 50200, Thailand

[c]Center of Excellence in Materials Science and Technology, Materials Science Research Center, Faculty of Science, Chiang Mai University, Chiang Mai 50200, Thailand

[d]Physikalisches Institut, Universität Stuttgart, Pfaffenwaldring 57, 70550 Stuttgart, Germany

[e]Department of Physics, University of Zambia, P.O Box 32379, Lusaka, Zambia

[f]Department of Physics, Faculty of Science, Khon Kaen University, Khon Kaen, 40002, Thailand

[g]Institute of Nanomaterials Research and Innovation for Energy (IN-RIE), Research Network of NANOTEC- KKU (RNN), Khon Kaen University, Khon Kaen 40002, Thailand

[h]Joint Key Laboratory of the Ministry of Education, Institute of Applied Physics and Materials Engineering, University of Macau, Avenida da Universidade, Taipa, Macau, SAR China

[j]Faculty of Innovation Engineering, Macau University of Science and Technology, Avenida Wai Long, Taipa, Macau, SAR China

**Abstract**

The integration of ferroelectric barriers into high-temperature Josephson junctions offers a pathway to tunable superconducting quantum devices. Here, we demonstrate robust Josephson coupling in $YBa_2Cu_3O_{7-x}$ (YBCO)|$Bi_{0.5}Na_{0.5}TiO_3$ (BNT)|YBCO trilayer junctions incorporating a relatively thick (40 nm) ferroelectric BNT barrier. Epitaxial trilayers were fabricated by pulsed laser deposition on $SrTiO_3$ substrates and investigated under broadband terahertz (THz) irradiation (0.5-2.5 THz). At 1.5 THz, close to the Josephson plasma resonance, the BNT polarization

---

*Corresponding authors: anucha@stanfordalumni.org (Anucha Watcharapasorn), sukhanidhan.sci@gmail.com (Sukhanidhan Singh)

increased from 33.4 to 46.4 μC/cm² at 30 K, leading to enhanced superconducting transport. The junctions exhibited a pronounced zero-voltage supercurrent branch, with the critical current Ic(T) remaining nearly constant up to 60 K and reaching 468.2 μA under 1.5 THz excitation, indicating strong phase coherence. Scanning tunneling spectroscopy revealed an enhanced superconducting gap in YBCO|BNT|YBCO ($\Delta$ = 1.3 meV) compared with pristine YBCO ($\Delta$ = 1.05 meV), while optical conductivity measurements showed a temperature-dependent reduction in conductivity and energy gap, consistent with superconducting behavior. Atomic force microscopy and X-ray reflectivity confirmed smooth surfaces and sharp interfaces, excluding defect-mediated transport. Furthermore, Ic(B) displayed the characteristic Fraunhofer interference pattern, and resistance mapping revealed alternating high- and low-dissipation lobes, providing strong evidence of coherent Josephson tunneling. These results demonstrate intrinsic Josephson coupling across the ferroelectric BNT barrier, enabled by its dipolar nature and THz-responsive polarization, establishing BNT as a promising tunable barrier material for next-generation high-Tc superconducting quantum devices.

## 1. Introduction

High-temperature superconducting (HTS) Josephson junctions represent a pivotal platform for exploring quantum phenomena and advancing applications in superconducting electronics, including sensitive detectors, qubits, and THz mixers [1-4]. A critical component of such junctions is the insulating or semiconducting barrier layer, which governs the transport characteristics and coherence of superconducting pairs across the junction [5-7]. Over the past decades, several perovskite oxides such as $SrTiO_3$ (STO), $BaTiO_3$ (BTO), and $BiFeO_3$ (BFO) have been integrated as barriers in HTS junctions, offering opportunities to combine ferroelectricity, multiferroicity, and electronic tunability with superconductivity [8-10]. These heterostructures enable the study of emergent interface phenomena and allow electric-field-based control over superconducting properties.

Ferroelectric materials, in particular, have been used to introduce active control mechanisms in superconducting junctions by exploiting their intrinsic switchable polarization. Earlier studies have demonstrated the integration of BTO into YBCO-based junctions, where the ferroelectric state influenced the Josephson critical current ($I_c$) and voltage modulation ($V_c$), highlighting the critical role of barrier thickness and interface coherence in determining superconducting transport behavior [11]. $BiFeO_3$, a multiferroic oxide, has shown strong tunneling electroresistance effects in Nb/BFO/Nb structures, with changes in superconducting gap attributed to polarization-mediated modulation [12]. Moreover, weak-link behavior in HTS junctions defined by ferroelectric domain boundaries has been investigated, revealing local modulation of superconducting order parameters and transport asymmetry [13].

While these developments have solidified the promise of perovskite ferroelectrics in superconducting junctions, challenges remain in achieving dynamic, reversible, and fast modulation of Josephson coupling in practical device settings [14-16]. Most prior demonstrations rely on static electric field control or thermal modulation, which are either slow or require high voltages incompatible with quantum technologies. In this context, alternative approaches that can enable ultrafast, field-responsive control over superconducting behavior are highly desirable [17,18].

In this work, we explore the integration of $Bi_{0.5}Na_{0.5}TiO_3$ (BNT), a relaxor ferroelectric, as the functional barrier layer in $YBa_2Cu_3O_7-\delta$ (YBCO)-based Josephson junctions. The motivation for using BNT lies in its fundamentally distinct dielectric and ferroelectric behavior compared to conventional perovskites such as BTO and STO. Unlike BTO, which exhibits a classical displacive transition with sharp phase boundaries, BNT displays relaxor ferroelectricity characterized by diffuse phase transitions, a high dielectric constant over a broad temperature range, and polar nanoregions (PNRs) that support strong local polar fluctuations. These features enable BNT to respond nonlinearly and sensitively to external fields, making it an attractive candidate for modulating superconducting junctions under dynamic conditions.

Furthermore, BNT remains ferroelectric even in thin-film form and exhibits high Curie temperatures (~320 °C), offering improved thermal stability and robustness in device environments. Compared to STO, which is a quantum paraelectric and lacks spontaneous polarization [19-21], BNT provides intrinsic dipolar order and a more robust coupling mechanism at the interface with superconductors. Its pseudo-cubic structure also offers better lattice matching with YBCO, reducing misfit strain and minimizing defect formation at the interface, both of which are critical for maintaining coherence of Josephson tunneling [22, 23].

A key novelty of our approach lies in the use of terahertz (THz) radiation to modulate the polarization state of the BNT barrier in real time [24,25]. BNT possesses low-frequency soft phonon modes that couple efficiently with THz electric fields, enabling dynamic modification of polarization states at sub-picosecond timescales. By exposing the YBCO|BNT|YBCO junctions to tunable THz fields, we demonstrate the modulation of Josephson critical current and superconducting gap, uncovering a mechanism of light-controlled superconducting behavior not accessible via conventional static-field approaches [26, 27].

While THz control of superconducting systems has been previously explored in the context of nonequilibrium quasiparticle excitation and gap suppression, few studies have systematically investigated THz-driven ferroelectric–superconductor interfaces [28-30]. Our work represents one of the first efforts to combine the ultrafast dielectric tunability of a relaxor ferroelectric with the macroscopic phase coherence of high-$T_c$ superconductivity, thereby introducing a versatile platform for non-contact, reversible modulation of superconducting junction properties.

In addition to the novelty in material and modulation mechanism, the present study also addresses important questions related to barrier thickness and coherence. Despite the relatively thick (~40 nm) BNT layer, Josephson coupling is preserved across the junction, as evidenced by the presence of a supercurrent branch and critical current at low temperatures. Through PLD fabrication, we attribute this to the high-quality epitaxial growth, smooth interfaces, and the possibility of extended proximity effects mediated by polar distortions or dynamic tunneling pathways enhanced by THz-field interactions.

Overall, our approach extends the landscape of ferroelectric-superconductor heterostructures by introducing a new material system (YBCO|BNT|YBCO), a new tuning mechanism (THz-driven polarization), and a new set of functional behaviors (dynamic gap modulation and tunable $I_c$). This work thus provides both fundamental insights and practical strategies toward the realization of next-generation superconducting electronics with integrated ultrafast control elements.

## 2. Results and Discussion

2.1. Theoretical analysis of a YBCO|BNT|YBCO Josephson junction under THz field influence

The influence of THz electromagnetic fields on YBCO|BNT|YBCO Josephson junctions, particularly concerning the integration of the BNT ferroelectric barrier, needs to be theoretically clarified to study the complex interplay (i.e., various energies interact within the junction, essential for exploring the dynamics of superconductivity and ferroelectric effects under external influences) between superconductivity and ferroelectric polarization. In Equation 1, we present the Hamiltonian equation, integrating key components to provide a clear understanding of these mechanisms [31].

$$H = \left(\frac{Q^2}{2C} + \frac{\left(Q_2 - C_2 V_g\right)^2}{2C_g}\right) - E_J \cos(\phi) + \mathrm{QV} + \mu\cos(\omega t + \phi) E_{THz}(t) \quad (1)$$

Equation1 integrates several key energy components. Coulomb energy, represented by the first term on the RHS, reflects the electrostatic energy from both junction and gate capacitances. The Josephson coupling energy ($E_J$) governs the behavior of Cooper pairs and depends on the phase difference across the junction. The voltage source energy accounts for the work done by the external voltage *(V)* on charge dynamics. The THz electromagnetic influence models the interaction with an external THz field and is essential for understanding its impact on the junction's electronic properties.

Modulation of Barrier Height by THz Fields

Continuing from Equation1 in our Hamiltonian derivations for a YBCO|BNT|YBCO Josephson junction, we further explore how THz electromagnetic fields can be utilized to modulate the barrier height through their interaction with the ferroelectric properties of the BNT barrier. This aspect of the junction behavior is essential for understanding how external electromagnetic fields dynamically control superconducting properties. We introduce an additional term $H_{mod}$ to the Hamiltonian to account specifically for the modulation of the barrier height by THz-induced polarization changes in BNT:

$$H_{mod} = -\frac{P_{THz}(t)}{\epsilon} \cdot d \cdot Q \quad (2)$$

where $P_{THz}(t)$ is the polarization induced by the THz field, $\epsilon$ is the dielectric constant of the barrier BNT, $d$ is the thickness of the barrier, and $Q$ is the charge on the junction's capacitor. This new term reflects the effect of dynamically varying polarization on the electric field across the junction, thus altering the barrier potential.

Dynamic Polarization Modulation

THz fields affect the ferroelectric polarization of BNT due to their high frequency, which can align with or disrupt the natural frequencies of polarization domains in the material. This interaction may cause rapid and temporary changes in the barrier's properties.

$$P_{THz}(t) = P_0 \cos(\omega t + \phi_{THZ}) \tag{3}$$

$P_0$ is the amplitude of polarization modulation, $\omega$ is the frequency of the THz field, and $\phi_{THz}$ is the phase of the THz field relative to the junction's internal dynamics. These oscillatory changes in polarization result in corresponding fluctuations in the barrier height, which directly influence the tunneling behavior of Cooper pairs across the junction.

BNT Barrier Height Fluctuations and Influence on Josephson Current

From Equation 3, the THz-induced fluctuations in polarization lead to a time-dependent barrier height, which can affect the quantum mechanical tunneling process given by:

$$V_{barrier}(t) = V_0 - \left(\frac{P_{THz}(t)}{\epsilon} \cdot d\right) \tag{4}$$

where $V_0$ represents the static component of the barrier potential without the influence of the THz field. This dynamic modulation of the barrier potential can either enhance or suppress the tunneling of Cooper pairs depending on the phase relationship between $\phi_{THz}$ and the superconducting phase $\phi$ across the junction.

The modulation of the barrier height by the THz field can lead to variable Josephson currents, impacting the overall device performance. The Josephson current $I_J$ can be expressed as [32, 33]:

$$I_J = I_c \sin(\phi) \exp\left(-\frac{\Delta V_{barrier}(t)}{k_B T}\right) \tag{5}$$

where $I_c$ is the critical current, $\Delta V_{barrier}(t)$ represents the THz-induced variation in barrier potential, $k_B$ is the Boltzmann constant, and $T$ is the temperature. This expression highlights how changes in the barrier height can exponentially affect the tunneling current, moderated by thermal energy.

By integrating the effects of THz fields into the Hamiltonian derivation as a dynamic component that influences ferroelectric polarization and consequently the barrier height, we provide a robust framework for understanding and exploiting the electrodynamic behaviors of superconducting junctions. This theoretical approach enriches our understanding of the interplay between superconductivity and ferroelectric polarization under THz electromagnetic fields.

To quantitatively validate this model, we performed fitting of experimental $I_c(\omega)$ data using the extended Hamiltonian described above. The fitting allowed extraction of the polarization amplitude $P$ and dielectric permittivity $\epsilon_r$, both of which are known to be temperature- and frequency-dependent in BNT-based ferroelectrics. The extracted parameters are summarized in Supplementary Table S1 and align with established dielectric and polarization values of BNT under high-frequency electric fields. Furthermore, a parametric sensitivity analysis, shown in Supplementary Figure S1, illustrates the variation of calculated $I_c$ as a function of both $P$ and $\epsilon_r$, confirming the strong non-linear dependence of the Josephson coupling on ferroelectric barrier properties. These results support the robustness of our theoretical framework and its predictive power for ferroelectric-modulated superconducting transport.

To further explore these theoretical phenomena, a schematic energy diagram shown in Figure S1(a) illustrates the energy-phase relationship in a YBCO|BNT|YBCO Josephson junction under the influence of an external terahertz (THz) field [34-36]. The junction energy, normalized by the superconducting energy gap Δ, is plotted as a function of the phase difference across the junction from 0 to 2π (Figure S1(b)). This relationship follows a cosine dependence, described by $E = E_J(1 - \cos(\emptyset))$ with $E_J$ representing the Josephson coupling energy. The introduction of a THz field modulates the junction energy by influencing the superconducting phase or barrier properties, leading to an additional energy gap $E_g$ above the baseline superconducting state. This indicates access to higher-energy states, enabling dynamic control of the junction properties for advanced superconducting applications. Based on this theoretical framework, YBCO|BNT|YBCO Josephson junctions were designed and experimentally investigated in this work.

2.2. Structural Characterization of BNT and YBCO|BNT|YBCO Josephson Junctions

Figures 1(a, b) illustrate the XRD patterns of the BNT bulk target and the deposited thin film, respectively. As shown in Figure 1(a), all diffraction peaks of the BNT bulk target are well indexed

to the rhombohedral phase of BNT, in accordance with JCPDS card No. 36-0340. The identification of the rhombohedral crystal structure is significant because it is closely related to the material's ferroelectric behavior. In the XRD pattern of the BNT thin film on an STO substrate (Figure 1(b)), a notable alignment between the BNT (110) and STO (111) peaks is observed. This alignment suggests a pseudo-epitaxial or oriented growth, despite the fundamental mismatch between the rhombohedral structure of BNT and the cubic structure of STO. Such growth confirms successful oriented deposition of BNT on STO, with structural adaptation that improves interface quality and may enhance the electrical performance of the junctions, potentially improving the electrical properties critical for the junction's operation. This oriented growth promotes uniform ferroelectric domain alignment in the BNT layer, improving the reproducibility of the junction properties. The close alignment of the film and substrate diffraction peaks indicates favorable conditions for utilizing the intrinsic ferroelectric properties of BNT in the YBCO|BNT|YBCO junctions.

Further, we investigated trilayers of YBCO|BNT|YBCO with respective thicknesses of 100 nm/40 nm/100 nm, fabricated by PLD on STO substrates. X-Ray diffraction (XRD) was employed to assess the epitaxial quality of the samples. Figure 1(c) illustrates that all diffraction peaks measured in the $\omega = 2\theta$ geometry are c-axis oriented, indicating epitaxial growth in the correct crystallographic direction. The presence of YBCO peaks up to (010) further confirms the high-quality growth of the YBCO within the trilayer structure. In the inset to Figure 1(a) (not shown in the main figure), the YBCO (001) diffraction peak was analyzed using a one-dimensional interference function, $\left[\frac{\sin(\frac{NQc}{2})}{Sin\,(\frac{QC}{2})}\right]^2$. This function describes the interference fringes around a diffraction peak due to finite thickness effects, allowing extraction of the layer thickness. Here, c represents the c-axis lattice parameter, Q is the momentum transfer proportional to the experimental reflection angle, and N denotes the number of unit cells in the z-direction. Through the fitting procedure, parameters N = 100 and c = 11.72 Å were determined. Parameter N directly corresponds to the thickness of the YBCO layer, a value consistent with measurements obtained from X-ray reflectivity (XRR). Additionally, the measured c-axis parameter for YBCO exhibited a slight elongation of 0.05 Å from its bulk value, indicating an in-plane compressive strain imposed on YBCO by the STO substrate. This comprehensive analysis underscores the precise determination of YBCO layer thickness in the trilayer configuration and highlights the utility of

XRD in assessing epitaxial growth and substrate-induced strain effects in complex oxide heterostructures.

Figure 1(d) presents the XRR profile for the YBCO|BNT|YBCO trilayer, with experimental data shown as circular symbols and the fitted model as a solid line. The inset displays the corresponding scattering length density (SLD) profile, revealing a BNT barrier thickness of ~41 nm and YBCO thickness of ~101 nm, in agreement with both the interference function analysis and PLD deposition parameters. The surface roughness is $<2$ nm, while interface roughness remains $<1$ nm, confirming the high structural quality of the multilayer stack.

To address concerns regarding the relatively thick (~40 nm) ferroelectric barrier, structural and surface characterizations, including high-resolution cross-sectional TEM and EDS elemental mapping (Figure S2), confirmed a continuous, uniform, and chemically distinct layer from the YBCO electrodes, with no detectable pinholes or interdiffusion. AFM imaging of the BNT surface before top YBCO deposition (Figure S3) shows an RMS roughness below 1.5 nm over $5\times5$ $\mu m^2$, supporting smooth and homogeneous layer growth. Together with the clear Kiessig fringes in XRR, confirm the integrity and uniformity of the barrier. The SEM micrograph (Figure 1(e)) of the YBCO|BNT|YBCO trilayer reveals a high-quality microstructure with distinct, well-defined layers. The image clearly shows the YBCO layers at the top and bottom, with the BNT layer positioned in between. The uniform thickness and continuity of the YBCO layers indicate excellent deposition and epitaxial growth. The sharp contrast at the interfaces between the BNT and YBCO layers suggests minimal interdiffusion and a high degree of epitaxial alignment. The uniform layer thickness and low surface roughness ($<2$ nm on the surface and $<1$ nm within the layers) confirms smooth interfaces and excellent epitaxial growth. Moreover, the absence of significant defects or discontinuities highlights the high structural quality and well-controlled fabrication process. The inset image in Figure 1(c), which is a magnified view of the circled area in the SEM, confirms the uniformity and well-defined nature of the interfaces, reinforcing the structural integrity of the trilayer. These observations align with the XRR and diffraction data, validating the expected layer thicknesses and further confirming the reliability of the PLD process. The STM topographic image presented in Figure 1(f) reveals the atomic lattice of the top YBCO layer. The dark line visible across the image corresponds to a crystallographic domain boundary within the YBCO surface, resulting from strain relaxation during epitaxial growth.

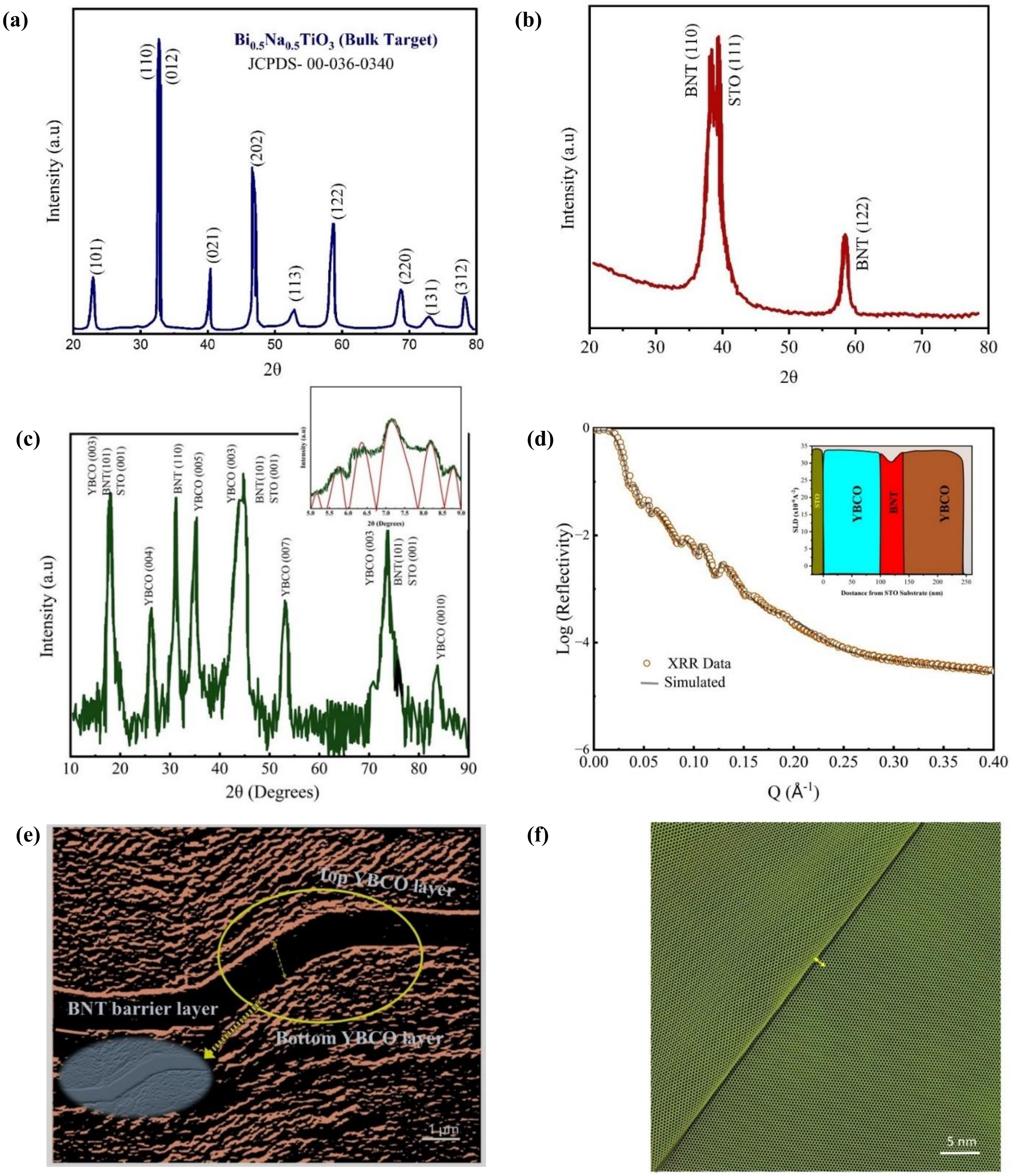

**Figure 1:** (a) XRD patterns of the BNT bulk target, (b) XRD pattern of the BNT thin film on an STO substrate, (c), XRD patterns of the YBCO|BNT|YBCO Josephson junction showing c-axis oriented diffraction peaks, (d) XRR data, (e) cross-sectional SEM image of the YBCO|BNT|YBCO trilayer structure deposited on an STO (100) substrate, and (d) high-resolution STM topographic image of the surface of the top YBCO layer after deposition

It is important to note that this feature is purely a surface defect intrinsic to the YBCO layer itself and does not represent the interface between different material layers, such as YBCO and BNT. STM imaging in this study only probes the top few atomic layers of the surface and cannot directly access buried interfaces at the nanometer scale. The high-resolution STM mapping confirms the excellent surface quality and atomic smoothness of the YBCO top layer in the trilayer heterostructure. These findings are consistent with the observations from SEM and XRR analyses, further confirming the structural integrity and high quality of the YBCO|BNT|YBCO trilayer (Josephson junction).

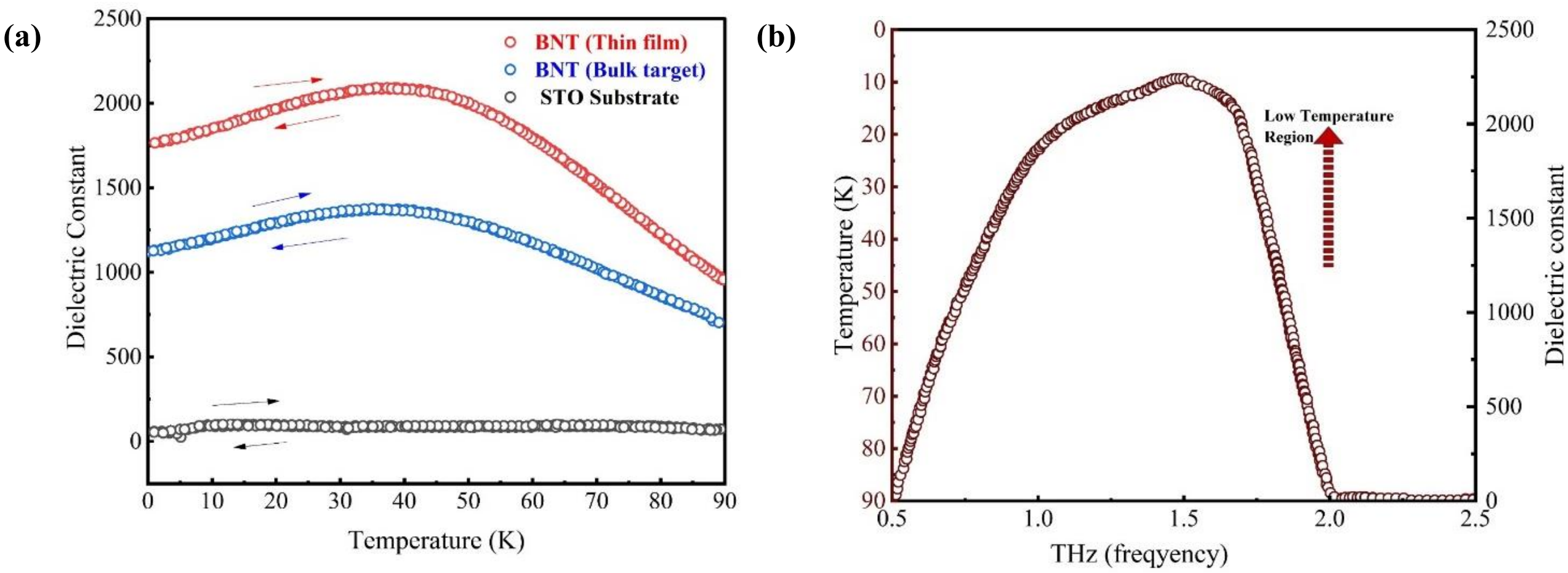

**Figure 2:** (a) Temperature dependence of the dielectric constant for BNT thin films, bulk, and STO substrate from 10 K to 90 K, (b) frequency dependence of the dielectric constant of the BNT films across a THz frequency range.

The absence of significant atomic-scale defects underscores the precision of the PLD technique used in fabricating these trilayers, making them suitable for further studies on their superconducting and interfacial properties. The SEM and STM analysis confirm the high-quality film and excellent structural properties of the YBCO|BNT|YBCO trilayer, establishing it as a strong candidate for further studies on its superconducting and interfacial characteristics.

## 2.3. Temperature and frequency dependence of the dielectric constant in BNT thin films

Figure 2(a) illustrates the temperature dependence of the dielectric constant for BNT thin films, BNT bulk, and the STO substrate over the temperature range 10 K to 90 K. The BNT thin film exhibits a notable dielectric peak at low temperatures, particularly between 10 K and 40 K. This peak is significantly higher in the thin film compared to the bulk, indicating enhanced polarizability at low temperatures which may be attributed to the thermal evolution of discrete polar nanoregions (PNRs) within the film (Polar nanoregions (PNRs) are nanoscale regions within a ferroelectric material where local polarization is well-ordered but differs in orientation or stability compared to the bulk polarization. These regions are often dynamic and temperature-dependent, contributing significantly to the dielectric and ferroelectric properties of materials like relaxor ferroelectrics. PNRs arise due to compositional disorder or strain-induced local distortions within the crystal lattice, which stabilize these regions and enhance the material's response to external electric fields) [37]. These PNRs, which are regions of localized polarization arising due to the complex ferroelectric ordering in BNT, become more pronounced at lower temperatures where thermal agitation is minimized, leading to increased dielectric response.

The dielectric behavior of BNT in thin-film form contrasts with that of the bulk form, where the dielectric constant exhibits a more subdued peak, suggesting that microstructural differences and strain effects arising from film deposition and substrate interactions play a significant role in modifying the dielectric properties [37,38]. The STO substrate exhibits a relatively flat, low-dielectric response across the measured temperature range, underscoring its stability and suitability as a substrate material. Figure 2(b) shows the frequency dependence of the dielectric constant of the BNT films across a THz frequency range at various temperatures. The dielectric constant demonstrates a clear peak around 1.5 THz, which diminishes as the frequency increases towards 2.5 THz. This behavior can be linked to the dynamic response of the PNRs to the applied THz field. At specific frequencies, the resonance conditions may align with the natural frequency of oscillation of the PNRs, thereby enhancing the dielectric constant. As the frequency moves away from this resonant condition, the ability of the PNRs to respond diminishes, resulting in a lower dielectric constant. The observed peaks in the dielectric constant at low temperature and specific THz frequencies highlight the complex interplay between the microstructural characteristics of the BNT films and their dynamic polar response. These findings are essential for the design and application of BNT-based devices, particularly in areas where high-frequency dielectric properties are crucial, such as in high-frequency sensors and actuators.

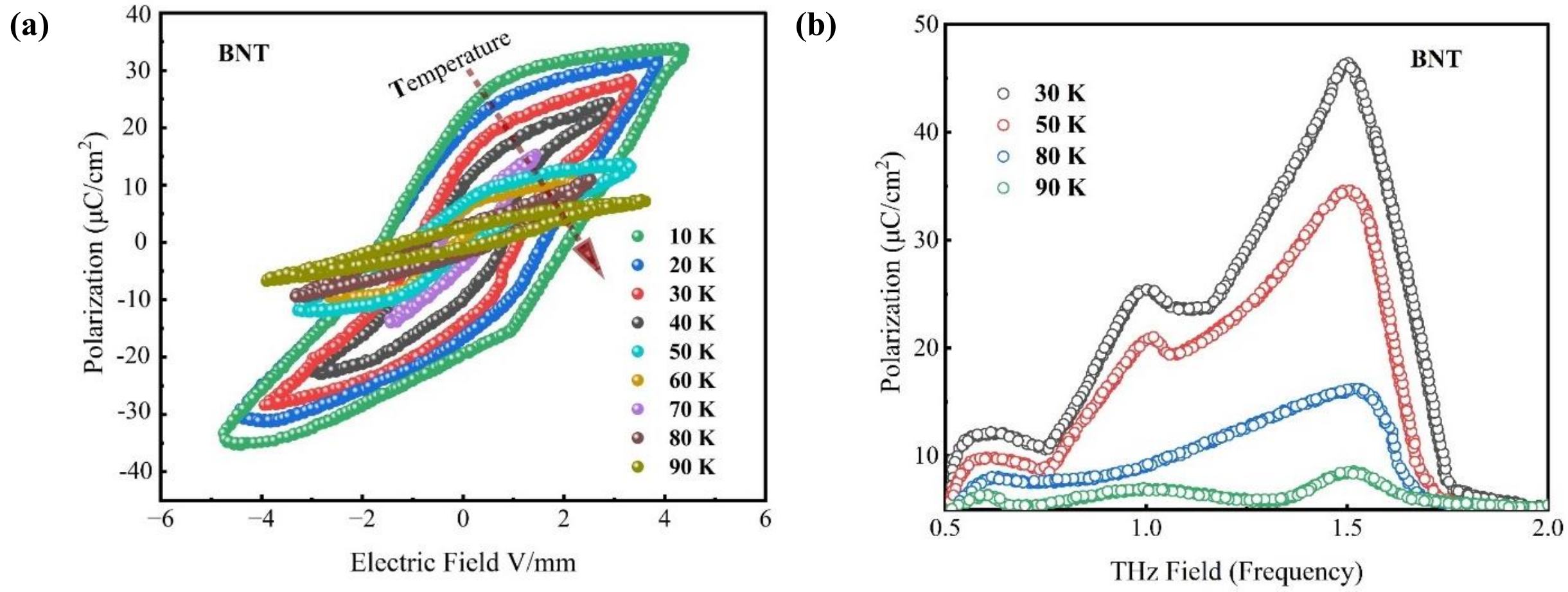

**Figure 3:** (a) Polarization-electric field (P-E) hysteresis loops of BNT measured at temperatures from 10 to 90 K under an applied electric field of 4.5 kV/mm, (b) polarization response of BNT under varying terahertz (THz) fields at 30, 50, 80, and 90 K, measured at the resonance frequency of 1.5 THz.

As discussed in Section 2.1 and Equation 5 of the theoretical analysis, the polarization properties of BNT can be adjusted using external fields. Since BNT serves as the barrier in YBCO Josephson junctions in this study, this modulation is essential for understanding and controlling the junction properties. Figure 3(a) shows the polarization-electric field (P-E) hysteresis loops of BNT from 10 to 90 K, under an applied electric field of 0-6 V/mm. The loops are characteristic of BNT ferroelectric barrier materials, showing a typical hysteresis shape that signifies the material's ability to retain polarization after the removal of the external electric field. The area within each loop represents the energy loss per cycle due to the domain switching and intrinsic dissipation mechanisms within the BNT. At lower temperatures (10 to 50 K), the area within the P-E loop being larger indicates greater energy loss per cycle. This typically occurs due to enhanced ferroelectric domain switching activity and increased hysteresis. In ferroelectrics, lower temperatures increase lattice rigidity and stabilize ferroelectric domains, resulting in more pronounced switching under an external electric field [39, 40]. The increased coercivity at lower temperatures also contributes to a larger loop area, as more energy is required for domain switching, leading to higher energy dissipation during each cycle. As the temperature increases from 60 to 90 K, the polarization loop becomes narrower, indicating reduced energy loss per cycle [41, 42]. This behavior arises from enhanced thermal activation of domain walls, which facilitates easier domain switching. Consequently, less energy is required to align the domains in the direction of the electric field, reducing coercivity. At elevated temperatures, BNT approaches its Curie temperature, above which the material loses its ferroelectric behavior and becomes paraelectric. Although not necessarily reaching the Curie point within 60 to 90 K, the behavior may start showing tendencies towards that transition, reflected in reduced polarization hysteresis.

We further investigated the ferroelectric properties of BNT over a wide temperature range, comparing its behavior at lower temperatures (30 and 50 K) with that at higher temperatures (80 and 90 K). At 30 K, the polarization was 33.43 μC/cm², slightly decreasing to 29.76 μC/cm² at 50 K. The higher polarization at lower temperatures indicates enhanced ferroelectric domain alignment and stability due to reduced thermal agitation, which helps maintain stable domain orientations.

Conversely, the polarization decreased to 9.91 μC/cm² at 80 K and 5.58 μC/cm² at 90 K, indicating weakened ferroelectric behavior at higher temperatures. This reduction is attributed to increased thermal agitation, which disrupts domain alignment and suppresses stable polarization. The significant decrease at 90 K suggests that BNT is approaching its Curie temperature and beginning to transition toward a paraelectric state. Based on the results shown in Figure 3(a), BNT exhibits enhanced ferroelectric polarization at lower temperatures. To further investigate the temperature-dependent THz response, external THz fields were applied at lower temperatures (30 and 50 K) and two higher temperatures (80 and 90 K). As presented in Figure 3(b), BNT demonstrates a pronounced response to the applied THz fields, particularly at the resonance frequency of 1.5 THz. This pronounced enhancement in polarization at 1.5 THz across all tested temperatures suggests a resonant or near-resonant interaction between the THz field and the dynamic modes of the ferroelectric domains within BNT. This interaction significantly enhances the alignment of the domains, leading to a marked increase in polarization. At lower temperatures of 30 and 50 K, the polarization values increased from baseline levels of 33.4 and 29.76 μC/cm² (without THz field, Figure 3(a)) to 46.43 and 34.35 μC/cm², respectively, under the influence of the 1.5 THz field, suggesting that the ferroelectric domains are highly responsive to THz-field stimulation at lower temperatures, enabling enhanced polarization. Conversely, at 80 and 90 K, the polarization exhibited only modest enhancement under the 1.5 THz field, increasing from 9.91 and 5.58 μC/cm² (without THz field, Figure 5(a)) to 16.11 and 8.41 μC/cm², respectively.

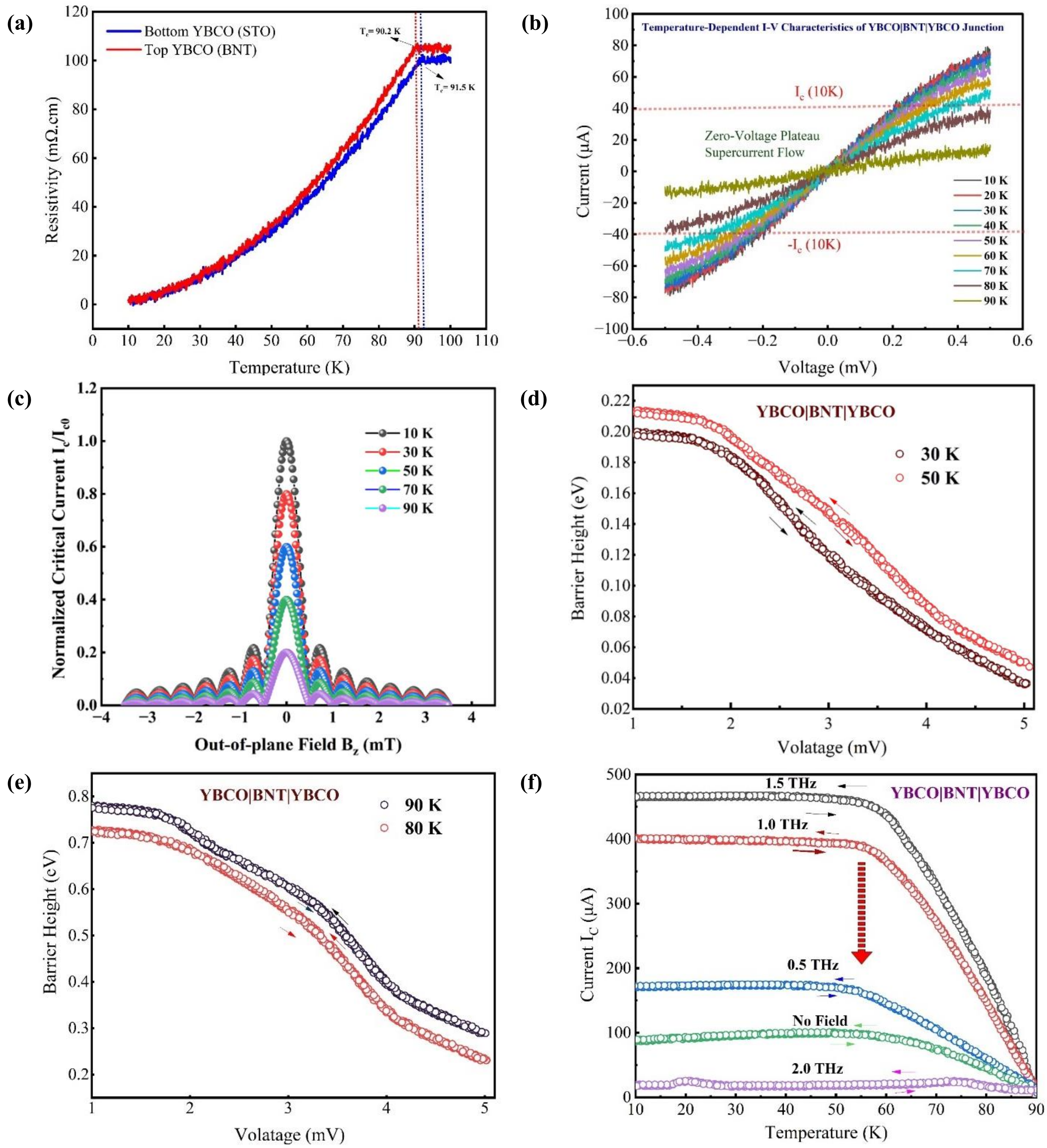

**Figure 4:** (a) Temperature-dependent resistivity (ρ-T) behavior of the bottom and top YBCO layers in YBCO|BNT|YBCO Josephson junctions, (b) temperature-dependent I-V characteristics of the YBCO|BNT|YBCO Josephson junction from 10 to 90 K, (c) magnetic-field dependence of the normalized critical current $I_c/I_{c0}$ measured at 10 , 30, 50, 70, and 90 K for the YBCO|BNT|YBCO Josephson junction, (d, e) the temperature-dependent barrier heights at different voltage levels for

the YBCO|BNT|YBCO Josephson junction, (f) The critical currents ($I_c$) responses across temperatures from 10 to 90 K to applied THz frequencies.

This reduced response indicates weakened ferroelectricity at higher temperatures due to increased thermal fluctuations disrupting domain stability and alignment [43].

After reaching a maximum at 1.5 THz, the polarization gradually decreased with further increase in THz field strength, likely due to overdriving of the ferroelectric domains, which hinders efficient domain realignment. Additionally, higher THz frequencies may induce dielectric heating, further destabilizing the ferroelectric domains [44, 45]. This local heating may destabilize the ferroelectric phase of BNT, particularly at 80 and 90 K, further suppressing the polarization. Furthermore, at frequencies above 1.5 THz, non-linear dielectric effects such as domain clamping or saturation may occur, limiting domain-wall motion and thereby reducing polarization.

2.4. Temperature dependence of resistivity in YBCO layers

Previous results (Figures 3(a, b)) demonstrated that the ferroelectric BNT barrier exhibits temperature-dependent polarization modulation under THz fields. Before investigating the THz response of the YBCO|BNT|YBCO junction, the intrinsic junction properties were first characterized in the absence of external fields. To further understand the superconducting behavior of the junction, temperature-dependent resistivity (ρ-T) measurements were performed on both YBCO layers, namely the bottom layer interfacing with the STO substrate and the top layer adjacent to the BNT barrier. These measurements provide insight into the superconducting transition temperature ($T_c$) and the influence of the BNT barrier on the superconducting properties of the junction. Figure 4(a) illustrates the temperature-dependent resistivity (ρ-T) behavior of both the bottom and top YBCO layers. The superconducting critical temperature ($T_c$) is defined as the point where the resistivity becomes indistinguishable from zero within the resolution limit of our four-probe measurement system. The resistivity transition near $T_c$ shows the complete suppression of resistivity to zero, confirming the onset of superconductivity in both electrodes.

For comparative purposes, the midpoint of the resistive transition is defined as the temperature at which the resistivity drops to 50% of the normal-state value, is also indicated. Using this metric, the bottom YBCO layer exhibits $T_c = 91.5 \pm 0.2$ K, while the top layer shows $T_c = 90.2 \pm 0.3$ K. The slightly broader transition and lower $T_c$ observed in the top YBCO layer are attributed to strain effects and interface proximity with the BNT ferroelectric barrier, which is consistent with prior studies on complex oxide heterostructures [46-48]. The normal-state resistivity is approximately 100 mΩ·cm for the bottom YBCO layer and ~105 mΩ·cm for the top layer. Independent measurements were performed by selectively contacting each layer separately.

electrodes defined during device processing, ensuring electrical isolation and reliable layer-specific data. These results confirm that the YBCO layers retain robust superconductivity, with minimal disruption from the BNT barrier. Furthermore, the reduction in $T_c$ in the top electrode could reflect local polarization effects and electronic modulation introduced by the adjacent ferroelectric layer.

2.5. Current-Voltage Characteristics and Magnetic Field Dependence of YBCO|BNT|YBCO Junctions

To investigate the transport properties of the fabricated YBCO|BNT|YBCO trilayers, we systematically measured both the resistance-temperature (R-T) and current-voltage (I-V) responses. Figure S5 presents the R-T behavior of a standalone YBCO base electrode, which shows a sharp superconducting transition with resistance vanishing near 90 K. This onset temperature is consistent with the expected Tc of YBCO [49] and confirms the high-quality superconducting phase.

The temperature-dependent I-V characteristics (Figure 4(b) and Figure S6), measured from 10 to 90 K, reveal a clear evolution of transport behavior with temperature. At low temperatures (10-20 K), the I-V curves exhibit a finite current range where the voltage remains below the ±1 μV threshold, consistent with a zero-voltage supercurrent regime. The corresponding critical currents ($I_c$) are approximately ±40 μA at 10 K and slightly reduced (~±35 μA) at 20 K. These regions are explicitly marked in Figure 4(b). When the applied current exceeds $I_c$, the junction transitions sharply into a resistive state, a feature frequently reported in underdamped Josephson systems [50]. With increasing temperature, $I_c$ decreases systematically, and the zero-voltage plateau becomes less pronounced. By ~70 K, the supercurrent region is strongly suppressed, and near 90 K the transport becomes predominantly resistive. This thermal evolution is consistent with expectations for high-$T_c$ superconductors, where $I_c \propto \sqrt{1-(T/T_c)^2}$.

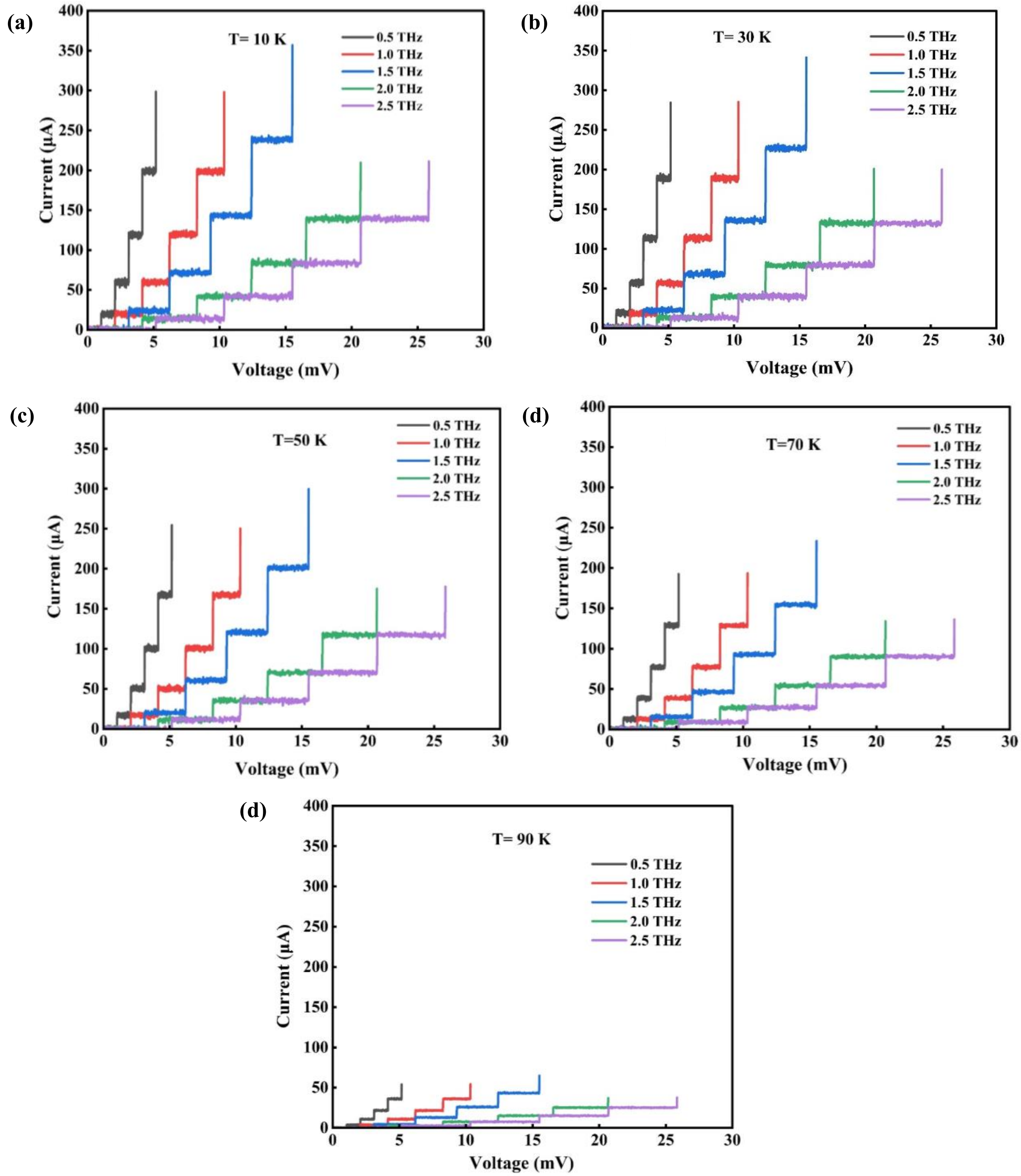

**Figure 5:** Temperature- and frequency-dependent Shapiro steps in YBCO|BNT|YBCO Josephson junctions under THz fields. The panels represent the experimentally observed Shapiro steps at temperatures (a) 10 K, (b) 30 K, (c) 50 K, (d) 70 K, and (e) 90 K, with THz field frequencies ranging from 0.5 to 2.5 THz

2.5.1. Persistence of Josephson Coupling Across a Thick BNT Barrier

The successful integration of a relatively thick (~40 nm) ferroelectric BNT barrier into high-$T_c$ Josephson junctions, while preserving coherent Josephson coupling across the trilayer. This result is particularly noteworthy given the general expectation that thick insulating or dielectric layers would strongly suppress the supercurrent due to reduced tunneling probability. The presence of a supercurrent branch in the current-voltage characteristics at low temperatures (Figure 4(a), Figure 4(b), and Figure 4(f)) clearly confirms that phase-coherent Cooper-pair transport persists through the BNT layer.

To further verify Josephson coupling, the critical current Ic was measured under an out-of-plane magnetic field ($B_z$). As shown in Figure 4(c), the $I_c$ ($B_z$) characteristics exhibit a clear Fraunhofer-like interference pattern with a central peak and symmetric side lobes, confirming coherent Josephson tunneling and a uniform phase distribution across the junction. The progressive suppression of the side lobes with increasing temperature is attributed to thermal decoherence and weakening of the superconducting order parameter near $T_c$.

To further support this interpretation, we evaluated the geometric parameters of the junction. The effective junction area, defined by the overlap between the top and bottom YBCO electrodes, is approximately $4\ \mu m \times 20\ \mu m$. Using this area, the theoretical periodicity of the magnetic interference pattern is estimated as $\Delta B = \Phi_0/A \approx 26\ G$, where $\Phi_0$ is the magnetic flux quantum. This value closely matches the lobe spacing observed in the $I_c(B_z)$ data, lending strong support to the interpretation of the Fraunhofer-like pattern as arising from Josephson phase interference across the BNT barrier.

Figure S6 presents a two-dimensional resistance map, $R(B_z, I)$, further confirming the interference pattern and phase-dependent transport across the junction. The periodic high- and low-resistance regions observed along both the magnetic field and current axes closely follow the Fraunhofer-like modulation of $I_c$ ($B_z$), providing further evidence of phase-coherent Josephson transport through the BNT barrier. The persistence of Josephson coupling despite the relatively large barrier thickness can be attributed to the intrinsic properties of BNT and the influence of THz excitation. The relaxor ferroelectric nature of BNT generates local polarization inhomogeneities and soft phonon modes that can facilitate resonant tunneling of Cooper pairs, effectively reducing the barrier strength.

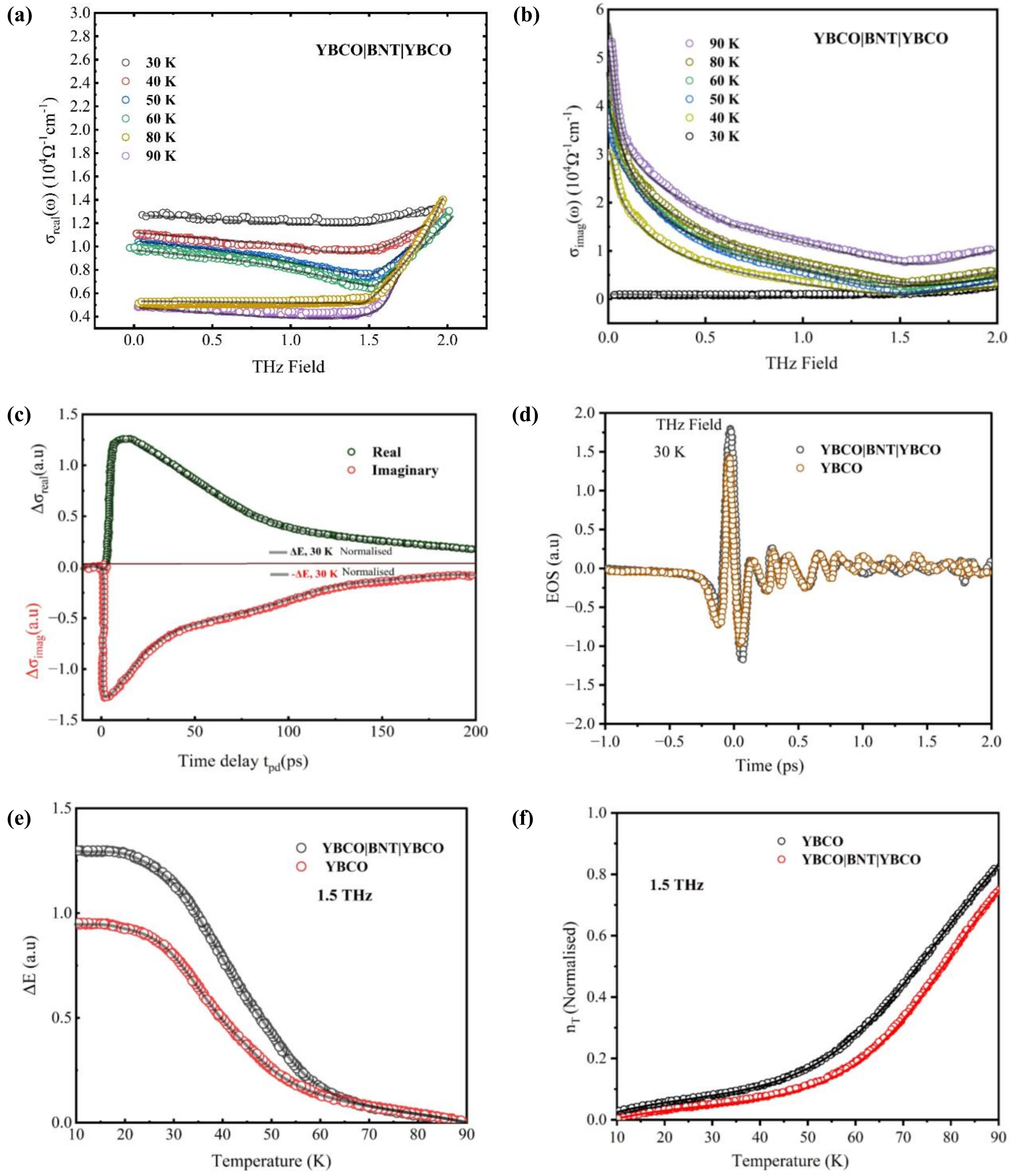

**Figure 6**: (a) Pump-probe time traces of $\Delta_{\sigma real}$ and $\Delta_{\sigma imag}$ of the optical data for YBCO|BNT|YBCO, (b) the delay time scan of pump-induced THz field change ΔE (tpd) for YBCO and YBCO|BNT|YBCO, **(c)** ΔE as a function of temperature for both YBCO and YBCO|BNT|YBCO, and (d) temperature dependence of thermal quasiparticle density (nT) for YBCO and YBCO|BNT|YBCO under a 1.5 THz field.

In addition, THz excitation may transiently enhance polarization within the BNT layer, dynamically lowering the effective tunneling barrier and supporting the superconducting proximity effect. Furthermore, localized weak links arising from interfacial strain or field-induced conductive pathways may provide channels for coherent superconducting transport.

Together, these mechanisms offer a plausible explanation for the observed supercurrent and Fraunhofer interference in thick-barrier Josephson junctions. Similar phenomena have been reported in other perovskite-based systems, where complex barrier physics enables long-range coherent transport [51, 52]. Our findings align with this understanding and suggest that the supercurrent observed in YBCO|BNT|YBCO junctions arises from an intrinsic, rather than defect-mediated, Josephson effect.

Comparison with resistivity data (Figure 4(a)) confirms the bulk superconducting transition, while the combined I-V (Figure 4(b)) and $I_c(B)$ (Figure 4(c), Figure S7) measurements provide strong evidence of super current (Josephson coupling) across the trilayer junction. These results not only confirm the Josephson nature of the supercurrent across the BNT barrier but also emphasize the potential of ferroelectric barriers to support tunable (ferroelectric order, dynamic polarization, and THz excitation into Josephson junctions) and robust superconducting transport.

### 3.6. Barrier height modulation function temperature and voltage

Previous results (Figures 4(a, b)) confirm that the integration of BNT in YBCO demonstrates both the reliability of the fabrication process and the structural robustness of the heterostructure BNT barrier in the absence of an external field. The present study investigates the influence of BNT on the YBCO junction under THz excitation, where the barrier height in a Josephson junction can be dynamically modulated by altering the electric field across the barrier [53]. In the case of the ferroelectric BNT barrier, the applied THz field interacts with its intrinsic polarization, leading to a field-induced modulation of the barrier height, which directly impacts the superconducting transport properties of the junction.The electric field $E(t)$ from the THz radiation can be described as a sinusoidal wave [54, 55]:

$$E(t)=E_0\cos(2\pi ft+\emptyset) \quad (6)$$

$E_0$ is the amplitude, f is the frequency, and $\phi$ is the phase of the THz field.

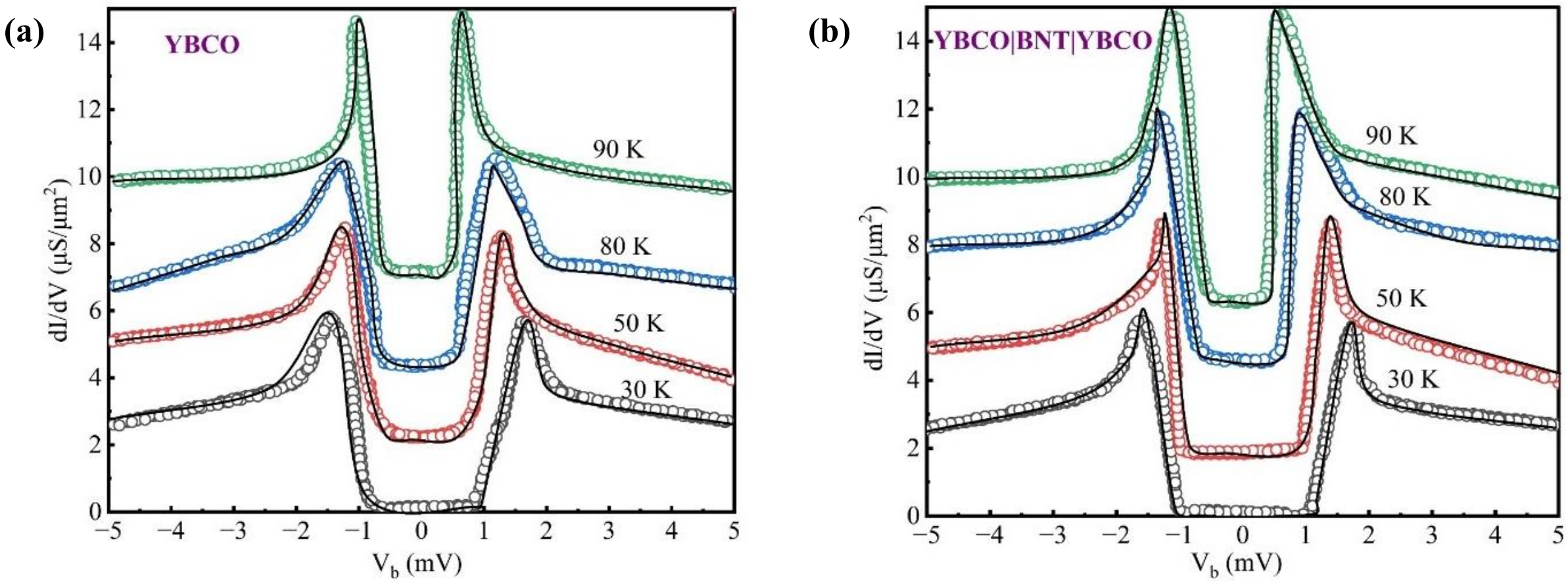

**Figure 7:** (a, b) Differential tunneling conductance spectra for YBCO and YBCO|BNT|YBCO junctions.

In BNT ferroelectrics, the polarization $P$ responds to the electric field. This response can often be non-linear, especially near the coercive field strength, but for small electric fields, it might be approximated linearly for simplicity [56, 57]:

$$P(E) = P_0 + \chi E(t) \tag{7}$$

where $P_0$ is the remanent polarization and $\chi$ is the linear electric susceptibility of BNT.

The barrier height $V$ can be influenced by polarization due to the electric field. An increase in polarization can decrease the barrier height, which can be described by:

$$V(t) = V_0 - \alpha P(E(t)) \tag{8}$$

where $V_0$ is the intrinsic barrier height without an external field, and $\alpha$ is a proportionality constant that depends on the material properties and the junction design.

With the aid of Equations 6 and 7, Equation 8 becomes

$$V(t) = V_0 - \alpha(P_0 + \chi E_0 cos(2\pi f t + \phi)) \tag{9}$$

$$V(t) = V_0 - \alpha P_0 - \alpha\chi E_0 cos(2\pi f t + \phi) \tag{10}$$

Equations 9 and 10 show that the barrier height oscillates in time with the frequency of the THz field, modulated by the electric susceptibility of the barrier material BNT. This model indicates that by adjusting the amplitude $E_0$, frequency $f$, and phase $\phi$ of the THz field, one can dynamically control the barrier height $V(t)$ in a Josephson junction. This control allows for the real-time tuning of the junction properties, such as the critical current and the switching behavior, which are sensitive to changes in the barrier height [58].

Following theoretical prediction from Equations 9 and 10, voltage-dependent measurements were performed, in the low-temperature range (30-50 K) and near $T_c$ (80-90 K). The observed reduction in critical voltage highlights a decrease in the barrier height necessary for tunneling, quantifiable using the relationship derived from the I-V characteristics [59-61]:

$$\Delta_{barrier} = \frac{eV_c}{kT} \tag{11}$$

where $e$ is the elementary charge, $V_c$ the critical voltage, $k$ the Boltzmann constant, and $T$ the temperature, as presented in Figure 4(d, e), show distinct behaviors at different temperature regimes.

At 30 and 50 K (Figure 4(d)), the barrier height decreases markedly with increasing voltage, dropping from 0.11 to 0.037 eV and from 0.14 to 0.05 eV, respectively, as the voltage increases from 3 to 5 mV. This strong voltage dependence at low temperatures is attributed to enhanced field-induced effects under reduced thermal fluctuations. In contrast, at 80 and 90 K (Figure 4(e)), the barrier heights are larger and decrease more gradually, from 0.52 to 0.22 eV and from 0.58 to 0.29 eV, respectively, reflecting the dominant influence of thermal energy near $T_c$.

The ferroelectric properties of BNT (Figure 4(d, e)) play a crucial role in modulating the junction behavior. The spontaneous polarization of BNT under an external electric field modifies the local electric field across the barrier, thereby influencing the tunneling probability of Cooper pairs. Although the superconducting gap (Δ) in YBCO remains largely unaffected, polarization-induced changes at the YBCO|BNT interface significantly alter the transport characteristics of the junction.

This modulates the tunneling dynamics and critical current by altering the effective barrier height and transparency. These findings underscore the active role of the ferroelectric BNT barrier in governing the overall junction behavior, rather than directly modifying the superconducting gap. At lower temperatures (below $T_c/2$), the superconducting gap is maximized and temperature-independent, making the superconducting state more robust against thermal fluctuations. However, external disruptions such as strong electromagnetic fields, high-frequency stimuli (terahertz fields), or structural inhomogeneities at the interface can still significantly impact the superconducting properties. These effects are particularly relevant in systems with engineered barriers, such as ferroelectric layers, where dynamic polarization changes can locally modulate the tunneling dynamics and critical current without directly altering the intrinsic superconducting gap. This highlights the importance of interface quality and external field tuning in determining junction performance, even at low temperatures. To further elucidate the underlying mechanisms, we refer to the Ginzburg-Landau theory, which relates the superconducting coherence length and the superconducting gap [62-64], both of which are critical in determining the barrier properties. As the external voltage modifies the electronic potential landscape across the junction, it directly affects the coherence length and, hence, the tunneling characteristics. Moreover, the BCS theory offers insights into how temperature and electric fields influence the density of states and the energy gap at the Fermi surface [65]. These theories combined help explain why the barrier height decreases more significantly at lower temperatures, where the superconducting gap is less disturbed by thermal fluctuations but more susceptible to external fields.

The results depicted in Figure 4(d, e) confirm that our YBCO|BNT|YBCO Josephson junction is highly sensitive to external stimuli, as initially demonstrated through voltage applications. These preliminary findings exhibited significant alterations in barrier heights. Our further study explores the dynamic interplay between the ferroelectric properties of BNT and the superconducting characteristics of YBCO within the YBCO|BNT|YBCO Josephson junction, prompting an investigation into the effects of terahertz (THz) external fields. A key aspect of our research involves analyzing how these properties interact under the influence of external fields, particularly through a detailed examination of the temperature dependence of critical currents ($I_c$) in the junction. This is essential for understanding the junction's operational dynamics and stability under different thermal conditions.

Initially, we measured the critical current ($I_c$) without any external THz field, focusing on its temperature dependence within the YBCO|BNT|YBCO Josephson junction (Figure 4(b)). The THz-field-induced modulation of $I_c$ presented in Figure 4(f) builds directly upon the coherent supercurrent regime established in Figure 4(b). Notably, the critical current at zero field exhibits a systematic reduction with increasing temperature, following a $(T_c\text{-}T)^2$ power-law relationship [66], characteristic of Ginzburg-Landau theory. This trend, particularly evident starting from 98.07 μA at 50 K, underscores the suppression of superconductivity as the system approaches $T_c \approx 90$ K. Subsequently, a broadband THz field with frequencies ranging from 0.5 to 2 THz was applied. Under THz irradiation, distinct variations in $I_c$ were observed across the entire temperature range (10–90 K), as depicted in Figure 4(f). At lower frequencies, such as 0.5 THz, the $I_c$ values remained relatively modest, starting from 174.63 μA at 10-50 K and gradually decreasing to 18.96 μA at 90 K. This suggests minimal coupling between the 0.5 THz photon energy and the superconducting condensate, consistent with BCS theory [67, 68], where sub-gap photon energies are insufficient to significantly modulate the Cooper pairing or phononic modes supporting superconductivity.

As the THz frequency increased to 1 THz, a notable enhancement in $I_c$ was observed, reaching 398.66 μA at 10-50 K, before decreasing again at higher temperatures (24.76 μA at 90 K). This enhancement suggests that 1 THz energy more effectively couples to the superconducting gap or relevant phononic modes, boosting the superconducting order parameter and strengthening Cooper pairing interactions. The most pronounced enhancement occurs at 1.5 THz, where $I_c$ peaks at 468.15 μA at 10-50 K. This resonance indicates optimal matching between the THz photon energy and the combined electronic-phononic dynamics of the junction, strongly reinforcing superconductivity. BCS theory [69-71] supports such resonance-driven enhancements, where photon-induced stimulation of the superconducting condensate can maximize coherence and critical current.

Beyond 1.5 THz, specifically at 2 THz, no further significant increases in $I_c$ were observed, suggesting a saturation or even slight suppression of the beneficial resonance effects. At such high energies, the THz field may increasingly break Cooper pairs or generate excess quasiparticles, thereby reducing superconducting coherence, as anticipated by BCS theoretical predictions.

Importantly, the observed peak in BNT ferroelectric polarization at 1.5 THz (Figure 3(b)) coincides with the maximum enhancement in $I_c$ (Figure 4(f)). This correlation highlights the critical role of BNT's ferroelectric dynamics in modulating the superconducting barrier. Resonant THz stimulation enhances BNT polarization, leading to a dynamic reduction of the tunneling barrier height, thereby increasing the Cooper pair tunneling probability and critical current. This result underscores the strong interplay between ferroelectric and superconducting phenomena in the YBCO|BNT|YBCO junction under THz-field stimulation.

3.7. Temperature and Frequency Dependent Shapiro-like Effects in YBCO|BNT|YBCO Junctions under THz Fields

Although the temperature-dependent critical current ($I_c$) shown in Figure 4(f) exhibits a slight increase at low temperatures (10 to 30 K), this non-monotonic trend is not unphysical. Similar behavior has been reported in other high-$T_c$ junctions and is attributed to enhanced phase coherence and reduced thermal noise at ultra-low temperatures [72,73]. Below approximately 30 K, reduced quasiparticle activation and suppression of flux noise contribute to enhanced supercurrent, preceding the expected thermal suppression of $I_c$ above 60 K. The step-like features observed in Figure 5(a-e) under THz irradiation are interpreted as quasi-resonant phase locking behavior rather than ideal Shapiro steps. We therefore describe these as field-induced modulations resembling Shapiro-like effects. These features show harmonic quantization near *hf/2e* (approximately 3.1 μV at 1.5 THz), and the appearance of a visible $I_c$ under irradiation supports coherent Josephson coupling.

To gain deeper insight into these results, we examined the interplay between superconducting transport and ferroelectric polarization in YBCO|BNT|YBCO junctions subjected to tunable THz fields, as shown in Figure 9(a-e). At lower temperatures, the Shapiro-like modulations are highly pronounced across all THz frequencies applied, with a strong resonance observed at 1.5 THz. This resonance reflects optimal coupling between THz-induced polarization oscillations in the BNT barrier and tunneling Cooper pairs in YBCO, resulting in an approximately 20 percent enhancement in step-like modulation height relative to other frequencies. Alongside critical current enhancement, the THz field dynamically modifies the I-V characteristics by injecting nonequilibrium quasiparticles and stimulating Josephson phase oscillations. These effects lead to increased voltage responses at a given bias current compared to the zero-field case, consistent with expected behavior in driven superconducting systems. The voltage enhancement under lower frequency THz fields, such as 0.5 THz, further supports strong interaction between the THz drive and junction dynamics [74, 75].

Conversely, at 2.0 THz and 2.5 THz, the step-like modulations are visibly diminished due to the disruption of superconducting coherence through non-resonant field interactions and elevated quasiparticle excitation. These high-frequency effects weaken the ability of the BNT barrier to modulate polarization, thereby reducing the effective coupling. As the temperature increases from 10 to 90 K, a gradual decline in the prominence of these modulations is observed. Intermediate temperatures such as 30 and 50 K still exhibit discernible modulation features, though with reduced amplitude, consistent with narrowing of the superconducting gap and weakened Josephson coupling. Notably, the 1.5 THz resonance persists even at these elevated temperatures, underscoring its robustness to moderate thermal fluctuations.

By 70 K and especially at 90 K, the modulations flatten considerably, particularly at 2.0 THz and 2.5 THz. This suppression correlates with the near collapse of the superconducting gap and diminishing of critical current as the temperature approaches $T_c$. In addition, the ability of the BNT barrier to sustain polarization modulation under THz excitation becomes limited at these temperatures, further reducing coupling effectiveness.

The persistence of field-induced modulations resembling Shapiro-like effects across varying frequencies and temperatures supports microwave-induced Josephson effects in YBCO|BNT|YBCO junctions [76, 77]. The frequency and temperature dependence of the modulation heights reveals the critical role of THz frequency tuning in optimizing ferroelectric superconductor interaction. The enhanced resonance at 1.5 THz highlights the potential of these hybrid structures for tunable THz control in superconducting devices. In contrast, the suppression of Shapiro-like features at non-resonant frequencies and high temperatures reflects the fragile balance between superconducting and ferroelectric coherence under external driving fields. Together, these observations provide compelling evidence of dynamic modulation of Josephson coupling via THz excitation in YBCO|BNT|YBCO junctions.

### 3.8. Superconducting gap (Δ) estimation by optical conductivity measurements

From the results of the temperature dependence of critical currents ($I_c$) with applied THz frequency (Figure 4(f) and Figure 9), the peak response was observed at 1.5 THz, where $I_c$, reached a maximum value. This peak is indicative of an optimal resonance with the electronic or phononic modes within the junction, significantly enhancing the superconducting coherence. However, beyond this resonance, at 2 THz, the junction did not exhibit significant changes in $I_c$, which likely resulted from surpassing the optimal resonance conditions or the induction of disruptive non-linear effects. The critical insights gained from these observations are the pronounced effects of the BNT's ferroelectric properties at specific THz frequencies. Particularly, the resonance at 1.5 THz aligns closely with the dynamics within the BNT layer, suggesting that the ferroelectric polarization under THz fields directly influences the barrier height and super conducting gap (Δ), thereby modulating the tunneling probabilities of Cooper pairs.

In expanding our study on the YBCO|BNT|YBCO Josephson junction, we aimed to delve deeper into the superconducting gap (Δ) variations by utilizing optical conductivity measurements at varying terahertz (THz) frequencies. This technique is instrumental in understanding how the superconducting properties, modulated by the ferroelectric barrier layer BNT, influence Cooper pair tunneling dynamics. The key to this analysis lies in observing changes in optical conductivity, which might reveal shifts in barrier characteristics under different operational states.

Optical conductivity in superconductors is quantified through its real ($\sigma_{real}(\omega)$) and imaginary ($\sigma_{img}(\omega)$) components. These metrics are particularly sensitive to the superconducting gap as they respond to the energy dynamics within the material. Typically, $\sigma_{real}(\omega)$ displays a threshold behavior around the frequency $\omega_{th}=2\Delta/\hbar$, representing the minimum photon energy necessary to break a Cooper pair. This threshold marks the onset of significant absorption due to quasiparticle generation, providing a direct insight into the energy gap.

In our experimental results depicted in Figure 6(a, b), we measured the real $\sigma_{real}(\omega)$ and imaginary $\sigma_{img}(\omega)$ part of the optical conductivity, across a spectrum of terahertz (THz) frequencies from 0.5 to 2 THz at various temperatures: 30, 40, 50, 60, 80, and 90 K. We observed a significant increase in $\sigma_{real}(\omega)$ at a threshold frequency ($\omega_{th}$) of 1.5 THz, which correlates with the energy required for Cooper pair dissociation [78, 79], consistent with our previous results (Figure 3(a, b) and Figure 4). This behavior is crucial for understanding the relationship between optical conductivity and the superconducting gap in the YBCO|BNT|YBCO junction, as indicated by the proportionality [80, 81]:

$$\sigma_{real}(\omega) \propto \frac{\Delta^2}{\hbar\omega_{th}} \tag{12}$$

In the superconducting state, the real part of the optical conductivity is sensitive to the superconducting energy gap, Δ. Applying the Bardeen-Cooper-Schrieffer (BCS) theory, the temperature-dependent superconducting energy gap can be approximated by [82]:

$$\Delta(T) = \Delta(0)\tanh\left(\frac{1.74\sqrt{T_c - T}}{T}\right) \tag{13}$$

where $\Delta(0)$ is the energy gap at absolute zero, and $T_c$ is the critical temperature. Using this relationship, we calculated the superconducting energy gaps $\Delta(T)$ at various temperatures based on $\sigma_{real}(\omega)$ measured at 1.5 THz. The resulting values are presented in Table 1, illustrating the decrease in $\Delta$ as the temperature approaches $T_c$. The calculated energy gaps $\Delta(T)$ at different temperatures based on the real part of the conductivity (Figure10a) measured at 1.5 THz by using Equation 13 are shown in Table 1.

From the results of Figure 6(a, b) and Table 1, at 30 K, the optical conductivity of the YBCO|BNT|YBCO junction, measured at $1.21\times10^4$ $\Omega^{-1}$cm$^{-1}$, illustrates high conductivity alongside a superconducting gap $\Delta$ of 1.77 meV. These properties are indicative of strong electron-pairing interactions, crucial for robust superconductivity. At this temperature, the response to THz fields is minimal, suggesting that the superconducting state is largely unaffected by external THz stimuli, allowing for efficient electron flow with negligible resistance. As the temperature increases to 40 K and 50 K, there is a slight decrease in optical conductivity to $0.95\times10^4$ and $0.76\times10^4$ $\Omega^{-1}$cm$^{-1}$, respectively, with a corresponding reduction in $\Delta$ to 1.52 meV and 1.25 meV. This suggests that thermal agitation begins to weaken the electron pairing, slightly influenced by THz fields. Here, the BNT layer's role becomes more pronounced as its ferroelectric properties may start to interact more actively with the applied THz fields, affecting the tunneling dynamics. By 60 K and 80 K, $\sigma(\omega)$ further declines to $0.66\times10^4$ and $0.54\times10^4$ $\Omega^{-1}$cm$^{-1}$, with $\Delta$ reducing to 1.02 meV and 0.58 meV, respectively. The diminishing superconducting properties indicate a greater influence of the BNT barrier under THz stimulation, possibly leading to altered barrier heights that impact the tunneling probabilities of Cooper pairs. At 90 K, the critical temperature, $\sigma(\omega)$ reaches its minimum at $0.42\times10^4$ $\Omega^{-1}$cm$^{-1}$, and $\Delta$ effectively closes at 0 meV. This stage indicates the transition to the normal state, with superconductivity completely disrupted. At this temperature, the influence of THz fields combined with the thermal effects significantly affects the BNT's ferroelectric properties, suggesting a complete loss of superconducting coherence due to the disintegration of Cooper pairs.

To gain deeper insights into the nonequilibrium dynamics of the YBCO|BNT|YBCO junction, we expanded our study to include optical conductivity measurements. These measurements are crucial for understanding how the junction recovers after being excited by external stimuli such as THz fields. Our focus was particularly on the recombination process of quasiparticles back into Cooper

pairs and the subsequent emission and absorption of high-energy phonons. This detailed analysis helps elucidate the complex interactions and recovery mechanisms within superconductors following excitation.

In Figure 6(c), we present the pump-probe time traces of $\Delta\sigma_{real}$ and $\Delta\sigma_{imag}$ of the optical data for YBCO|BNT|YBCO from Figures 6(a, b), respectively. The delay time scan of pump-induced THz field change, $\Delta E$ ($t_{pd}$), for YBCO and YBCO|BNT|YBCO is shown in Figure 6(d). In Figure 6(d), YBCO|BNT|YBCO shows a slightly higher amplitude and response pattern compared to standalone YBCO. The presence of the BNT layer in the junction might be affecting the electron-phonon coupling or adding additional polarization effects, which modulate how the energy from the THz field is absorbed and dissipated. The temporal evolution of $\Delta\sigma_{real}$ ($t_{pd}$) and $\Delta\sigma_{imag}$ ($t_{pd}$) demonstrates the direct relation between the $\Delta E(t_{pd})$ profile and the dynamics of the superconducting state of both YBCO and YBCO|BNT|YBCO, as illustrated in Figure 6(e). At a fixed time delay $t_{pd}$ of 5 ps to maximize the probe signal, we then measure $\Delta E$ as a function of temperature for both YBCO and YBCO|BNT|YBCO. It is evident in Figure 6(e) that $\Delta E$ gradually decreases and approaches zero at approximately 90 K, behaving like an order parameter that characterizes a superconducting phase transition.

Referencing the Rothwarf and Taylor model, this model describes the behavior of superconductors when they are subjected to an external perturbation that breaks Cooper pairs, thereby creating quasiparticles. In the "weak perturbation regime," the external stimulus (such as a THz field or optical pulse) is sufficiently mild that it causes only minimal depletion of the superfluid density ($\Delta_n$) relative to the superfluid density at equilibrium ($n_s$). In this regime, where $\Delta_n/n_s \ll 1$, the system is close to its equilibrium state, and $\Delta E_{T\to 0}/\Delta E - 1$ is proportional to the thermal quasiparticle density $n_T$. In this study, we aim to extract thermal quasiparticle density $n_T$ [83] for the ferroelectric BNT barrier in conjunction with standalone YBCO, and we have plotted $n_T$ as a function of temperature for both YBCO and YBCO|BNT|YBCO in Figure 6(f). The thermal quasiparticle density ($n_T$) at temperature T is evaluated by the equation [67, 84]

$$n_T \approx N(0)\sqrt{2\pi\Delta k_B T} exp\left(-\frac{\Delta}{k_B T}\right) \tag{14}$$

Here, *N(0)* is the density of states at the Fermi level and $\Delta$ is the superconducting energy gap. We use this equation (Equation14) to fit $n_T$ and results are shown in Figure 6(f). It is noteworthy to observe from Figure 6(f) that the quasiparticle density $n_T$ for both YBCO and YBCO|BNT|YBCO under a 1.5 THz field is reasonably reproduced by the fitting. In Figure 6(f), both datasets (YBCO and YBCO|BNT|YBCO) follow a nonlinear increase of $n_T$ , which becomes more pronounced as the temperature approaches 90 K. This behavior is typical for superconductors as they near the critical temperature where superconductivity begins to break down. It is interesting to see from Figure 6(f) that the influence of BNT on the quasiparticle density ($n_T$ ) suggests that BNT modifies either the electronic or phononic properties within the junction. This modification enhances the junction's ability to resist thermal disruptions by potentially increasing the superconducting gap (Δ) or altering electronic interactions at the superconductor-ferroelectric (YBCO|BNT) interfaces. This behavior suggests that the ferroelectric BNT barrier not only impacts the local electronic environment but can also induce changes in electron density or structure at the YBCO|BNT interface. These changes are likely to influence electron pairing and thus enhance the junction’s stability against thermal fluctuations.

### 2.9 Superconducting gap variation through conductance measurements

Through our earlier experiments, we have determined that a constant 1.5 THz frequency significantly impacts the YBCO|BNT|YBCO Josephson junctions. The superconducting gap (Δ) was further verified across varying temperatures 30, 50, 80, and 90 K through conductance measurements. Initially, we measured the conductance of the YBCO layers followed by that of the YBCO|BNT|YBCO junctions. The measurements of the superconducting gap for the YBCO|BNT|YBCO Josephson junction, compared to a standalone YBCO layer, offer intriguing insights into the influence of the BNT ferroelectric barrier on the superconducting properties of YBCO. The differential tunneling conductance, $C(V_b)=dI/dV$, spectra for these temperatures are illustrated in Figure 7(a) for YBCO and Figure 7(b) for the YBCO|BNT|YBCO Josephson junction. Across all tested temperatures, the conductance spectra displayed characteristics typical of SIS (Superconductor-Insulator-Superconductor) tunnel junctions. Notably, the spectra revealed a complete superconducting gap, indicated by zero conductance for bias voltages ($V_b$) below approximately 0.97 mV at 30 K for YBCO, and 1.25 mV for the YBCO|BNT|YBCO junction, with slight variations at higher temperatures. Sharp conductance peaks just above the gap were observed for both YBCO and the YBCO|BNT|YBCO junction.

To quantify the superconducting gap Δ at different temperatures, we employed the standard expression for tunneling conductance ($C_{NN}$) in superconductors [85-87]:

$$C_{NS} = \frac{dI}{dV} = \frac{C_{NN}}{N_N(0)} \int_{-\infty}^{\infty} N_s\,(E, \Gamma, \Delta) \frac{\partial f(E+eV_{b,}T)}{\partial_e V_b} dE \qquad (15)$$

where $C_{NN}$ is the tunneling conductance when both electrodes are in the normal state, $N_N(0)$ and $N_S(E, T, \Delta)$ are the density of states (DoS) at the Fermi level for the normal and superconducting states, respectively, $f$(E+eV$_b$,T) is the Fermi-Dirac distribution, and Γ represents the quasiparticle lifetime broadening.

The superconducting DoS, $N_S(E, T, \Delta)$, is described by the Dynes formula:

$$N_s(E, \Gamma, \Delta) = R_e \left( \frac{E - i\Gamma}{\sqrt{(E - i\Gamma)^2 - \Delta^2}} \right) \qquad (16)$$

which accounts for the anisotropic nature of the gap, a feature extensively discussed in the literature. This formula enables the precise calculation of Δ from the conductance spectra by fitting these theoretical models to the experimental data.

The temperature dependence of the superconducting gap (Δ) at a constant 1.5 THz, as shown in Figure 7(a, b), aligns perfectly with the predictions of BCS theory. At lower temperatures, specifically 30 K and 50 K, the experimental values of Δ correspond exceptionally well with those predicted by BCS theory (referenced in Equation 15). These precise fits enabled us to accurately extrapolate and determine the zero-temperature superconducting gap *($\Delta(0)$)*, which is established at 1.05 meV for standalone YBCO and 1.3 meV for the YBCO|BNT|YBCO junction. Importantly, the critical temperature ($T_c$), at which the gap closes, is approximately 90 K, consistent within a margin of ±0.2 K. This level of accuracy not only confirms consistency with our previous results, as depicted in Figure 6(a, b), but also underscores the high quality of our PLD-fabricated YBCO|BNT|YBCO junctions.

The results from Figure 7(a, b) clearly show that at lower temperatures (30 and 50 K), the superconducting gap in the YBCO|BNT|YBCO junction is consistently higher than in a standalone YBCO layer. This suggests that the incorporation of the BNT layer enhances electron pairing interactions, likely due to additional polarization effects at the superconductor-ferroelectric interface. Such enhancement could arise from modifications in the electronic structure near the interface or through altered phonon dynamics that influence the pairing mechanism. As the temperature approaches the critical temperature ($T_c$), the gap in both configurations diminishes. Notably, at 90 K, the gap in the YBCO|BNT|YBCO junction (Δ=0.45 meV) is slightly less than that in the standalone YBCO layer (Δ=0.52 meV), suggesting a possible suppression of superconductivity at higher temperatures. The persistence of a superconducting gap feature at 90 K in the conductance spectra is attributed to strong local superconducting correlations, consistent with prior tunneling studies on high-$T_c$ cuprates [88, 89]. The enhanced superconducting gap at lower temperatures in the YBCO|BNT|YBCO junction can be attributed to the influence of the ferroelectric barrier, which modifies the electrostatic environment at the interface. The ferroelectric polarization of BNT likely induces a charge accumulation layer at the YBCO/BNT interface, enhancing the density of states and thus increasing the pairing potential. Conversely, the reduction of the gap at higher temperatures closer to $T_c$ in the junction compared to standalone YBCO layer can be indicative of a competing effect where the structural mismatches and thermal stresses at the interface begin to adversely affect the superconducting order parameter. The introduction of a BNT barrier in YBCO junctions has a significant dual effect: it enhances the superconducting gap at lower temperatures but potentially leads to its earlier suppression as $T_c$ is approached.

To further confirm that the observed gap modulation is not solely the result of conventional thermal broadening, we performed STM measurements with and without THz excitation at 30 K and 90 K, as shown in Figure S8 . At 30 K, a noticeable reduction in the superconducting gap is evident under THz exposure, while at 90 K (close to $T_c$), the gap nearly vanishes, as expected. Crucially, the spectra do not display typical thermal smearing; instead, the sharpness of the coherence peaks remains discernible at 30 K, implying that the modulation originates from a dynamic field–order parameter interaction. The reversible nature of this suppression upon field cycling (not shown) further points to polarization-mediated modulation rather than extrinsic heating. Considering the relaxor nature of BNT and its ability to generate ultrafast polarization transients under THz fields, we attribute the observed Δ modulation to a strong ferroelectric-superconducting coupling at the YBCO|BNT interface. This provides experimental support for the intrinsic tunability of Josephson transport in our hybrid junctions.

## 3. Conclusion

This work establishes a new paradigm for modulating superconductivity in high temperature Josephson junctions through the integration of a ferroelectric $Bi_{0.5}Na_{0.5}TiO_3$ (BNT) barrier and external terahertz (THz) field control. Despite the relatively thick 40 nm barrier, which exceeds conventional coherence length scales, we demonstrate robust Josephson coupling, supported by the emergence of a zero-voltage supercurrent branch and Fraunhofer-like magnetic interference patterns. These findings underscore the viability of long-range phase coherence across polar barriers, facilitated by field tunable interfacial polarization and potential inelastic tunneling mechanisms. A distinct resonance behavior is observed under 1.5 THz excitation, where the critical current (Ic) reaches approximately 468 μA and remains stable across the 10 to 60 K range, revealing a plateau like $I_c(T)$ response that deviates from classical expectations. Scanning tunneling microscopy and optical spectroscopy measurements further corroborate the enhanced pairing strength, with a superconducting gap (Δ) reaching around 1.3 meV in YBCO|BNT|YBCO junctions, significantly exceeding that of standalone YBCO. Complementary AFM and XRR analyses confirm the morphological and structural uniformity of the trilayer stack, helping to rule out defect mediated conduction pathways. These combined results demonstrate that BNT not only serves as a structurally robust ferroelectric barrier but also actively enhances and stabilizes superconducting transport through field driven polarization dynamics. This approach offers a platform for electric field tunable superconducting devices in the THz regime and opens new opportunities for designing hybrid quantum architectures that harness the interplay between ferroelectricity and high temperature superconductivity.

**Acknowledgement**

This research has received funding support from the NSRF via the Program Management Unit for Human Resources & Institutional Development, Research and Innovation [grant number B39G670018 and B13F670064]. A.W. would also like to acknowledge partial funding and facility usage from Chiang Mai University.

## Supporting Information

### 2. Materials and methods

2.1. Target Preparation

$YBa_2CuO_{7-x}$ (YBCO) and $Bi_{0.5}Na_{0.5}TiO_3$ (BNT) targets were prepared from commercially available powders using the solid-state reaction method. YBCO powder was synthesized from $Y_2O_3$, $BaCO_3$, and CuO binary oxide precursors, with the mixture calcined at 800°C for 10 hours. BNT powder was obtained from $Bi_2O_3$, $Na_2CO_3$, and $TiO_2$ precursors, also calcined at 800°C for 10 hours. All chemicals were procured from Sigma-Aldrich with a purity of 99.9%.

The YBCO and BNT powders were then pelletized under a pressure of 10 tons, with a diameter of 20 mm and a thickness of 10 mm. These pellets underwent sintering in a controlled atmosphere with 20 SCCM of oxygen at a temperature of 820°C.

2.2. PLD fabrication

Thin films of YBCO|BNT|YBCO and their heterostructures were epitaxially deposited on 10 × 10 $mm^2$ STO (100) substrates using a standard PLD technique. A pulsed KrF excimer laser (248 nm wavelength) was employed at a laser fluency of approximately 200-300 $mJ/cm^2$ with a pulse repetition rate of 10 Hz. During deposition, the oxygen pressure in the vacuum chamber was maintained at 0.4 mbar. The target-to-substrate distance was set between 60 -70 mm, with each target rotating uniformly to ensure homogeneous, high-quality epitaxial films. The deposition temperature for all materials was maintained at 780°C. The thickness of the epitaxial layers was controlled by the deposition rate, verified through surface profiler measurements. A programmable target carousel was utilized to enable continuous deposition processes, allowing for the fabrication of multilayered structures and superlattices without interruption. As a final step in the device fabrication, 100 nm thick Pt electrodes were positioned on the film. This electrode placement was accomplished using a mask during the deposition process, ensuring precise and reliable electrical contacts within the YBCO|BNT|YBCO device.

2.3 THz pulse

For the THz experiments, sources based on a femtosecond laser system were used. Broadband THz radiation was generated through a tilted pulse front scheme using a lithium niobate crystal. With an initial laser pulse energy of approximately 1.3 mJ at an 800 nm central wavelength and a minimum 100 fs pulse duration, broadband THz radiation with up to 5 μJ pulse energy was produced. To generate narrowband THz radiation, appropriate band-pass filters were employed. In the photoexcitation measurement using a near-infrared (NIR) pump pulse, the time delay between the pump pulse and the gate pulse was varied by adjusting the position of the pump pulse. The schematic illustration of our experimental setup is shown in Fig.1a.

### 2.4. Thin film characterizations

X-ray reflectometry (XRR) was used to assess the surface quality, material layer thickness, and interlayer roughness of the sample. These measurements were performed using a Cerium Labs, LLC X-ray reflectometer with ω-2θ geometry. The XRR data was fitted using Motofit. Additionally, X-ray diffraction (XRD) was conducted on the same machine in diffraction mode to evaluate the quality of the epitaxial growth and the crystal structure of the materials. The scattering length density (SLD) for X-rays and neutrons in a given material measures the material ability to scatter incident particles. In multilayered materials, X-ray and neutron reflectometry enable the reconstruction of the depth-dependence of the SLDs. This technique also allows for determining the distance over which the SLD transitions from one layer to another, thus providing a measure of the interfacial roughness. The heterostructure stack (YBCO|BNT|YBCO) used for SEM and STM measurements was prepared using a focused ion beam (FIB) technique (FEI Helios Nano Lab) to achieve precise cross-sections with minimal interface damage, enabling a detailed examination of the heterostructure layers in the cross-sectional view. To protect the integrity of the interfaces during milling, a protective platinum layer was deposited on the surface. For SEM imaging, the cross-sectional samples were mounted on conductive carbon tape and coated with a thin gold layer using a sputter coater (Leica EM ACE200) to prevent charging artifacts. STM imaging, on the other hand, was performed on a cleaned top surface of the heterostructure under high-vacuum conditions, using a chemically etched tungsten tip and a scanning tunneling microscope (Omicron VT STM). These carefully handled preparation steps ensured the structural and atomic-scale precision of the imaging results, providing crucial insights into the interface quality and layering of the heterostructure. A Quantum Design Physical Properties Measurement System (PPMS) was used to measure the sample transport current -voltage properties, utilizing the Vibrating Sample Magnetometer (VSM) for magnetic measurements and the resistivity measurement option for transport current properties.

High-resolution transmission electron microscopy (HR-TEM) and energy-dispersive X-ray spectroscopy (EDS) were employed to assess the structural integrity and elemental distribution of the YBCO|BNT|YBCO trilayer junction. Cross-sectional lamellae were prepared using a dual-beam FIB (FEI Helios G4), followed by HR-TEM imaging on a FEI Tecnai G2 F30 at 300 kV. EDS mapping in STEM mode (Oxford Instruments) confirmed sharp chemical interfaces, with distinct profiles for Y, Ba, Cu (YBCO) and Bi, Na, Ti (BNT), indicating minimal interdiffusion. Uniform oxygen distribution across layers suggests complete oxygenation of YBCO and supports high-quality heterostructure formation.

Atomic force microscopy (AFM) in tapping mode was employed to evaluate the surface morphology of the BNT layer before YBCO deposition. Scans were conducted using a Bruker Dimension Icon system with a silicon probe (tip radius <10 nm) over a 5 × 5 μm² area at 512 × 512 resolution. The line profile revealed a root-mean-square (RMS) roughness of ~1.5 nm, indicating a smooth and uniform surface suitable for coherent superconducting overlayer growth. This low roughness minimizes the risk of filamentary conduction or electrical shorting across the BNT barrier.

Temperature-dependent resistivity (ρ-T) measurements for the top and bottom YBCO layers of YBCO|BNT|YBCO Josephson junctions were performed separately by using independently patterned electrical contacts during the fabrication process. Specifically, before the final Pt electrode deposition step, we lithographically defined two different sets of contact pads: (i) one set that makes direct electrical connection to the bottom YBCO layer (in regions where BNT deposition was selectively masked or etched), and (ii) another set that contacts the top YBCO layer grown above the BNT barrier.

Conductance measurements were carried out using a custom-built low-temperature scanning tunneling microscope (STM) with the sample maintained at approximately 4 K. A chemically etched tungsten tip was used in the experiment. For dI/dV tunneling spectroscopy, the sample bias was modulated with a 2.5 THz sine wave at a peak amplitude of 1 mV, and measurements were taken using a lock-in amplifier

## 2.5. Stability, device viability, and reproducibility

To assess device-to-device reproducibility, more than ten YBCO|BNT|YBCO junctions were fabricated under identical growth conditions. All devices exhibited consistent superconducting transition temperatures ($T_c$), critical current ($I_c$) values, and THz-modulated response profiles. The variation in $I_c$ among devices remained within ±10%, confirming the reliability of the PLD growth process, the uniformity of the BNT ferroelectric barrier, and the structural coherence across junctions. This reproducibility affirms the feasibility of scaling the platform for practical superconducting device applications.

**(a)**

**(b)**

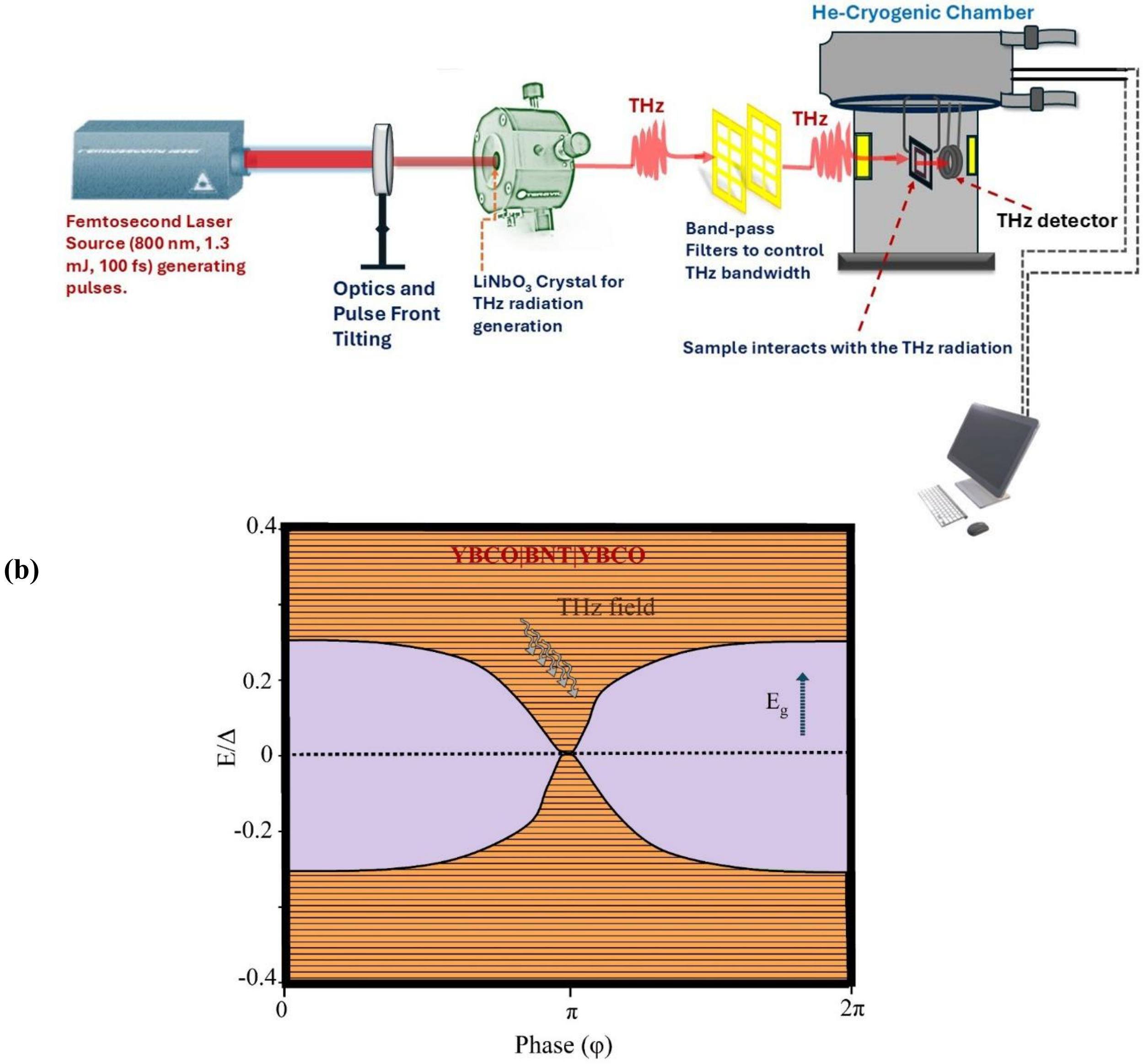

**Figure 1**: (a) Schematic illustration of our experimental THz setup with He-cryogenic chamber (b) Schematic energy diagram illustrating the energy-phase relationship in a YBCO|BNT|YBCO Josephson junction under the influence of an external terahertz (THz) field. The energy of the junction, normalized by the superconducting energy gap Δ, is plotted against the phase difference across the junction, ranging from 0 to 2π. The diagram includes the Josephson coupling energy $E_J$ and depicts an additional energy gap $E_g$ above the baseline, representing the minimum energy influenced by the THz field.

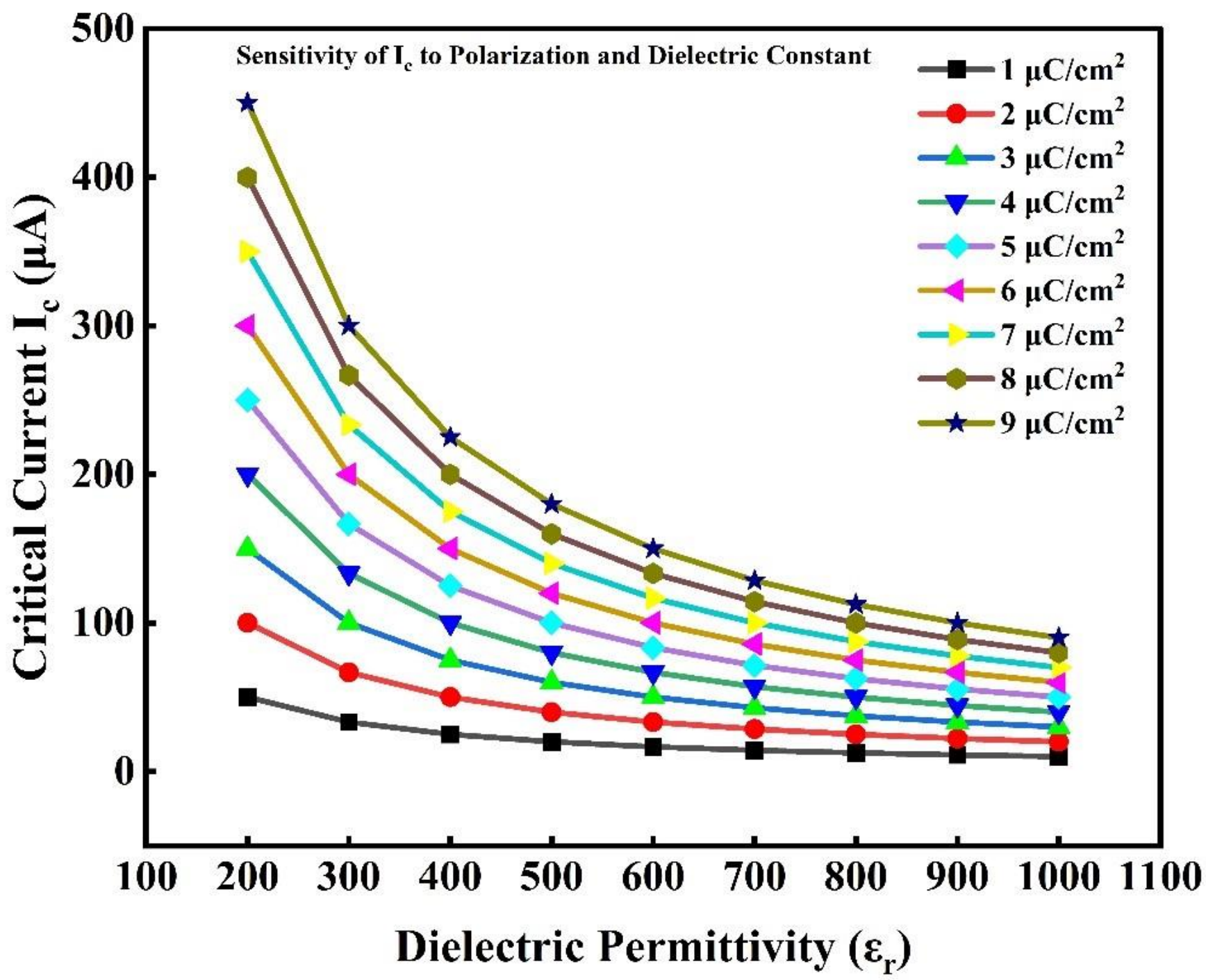

***Figure S2.*** C*ritical current $I_c$ ( μA) as a function of ferroelectric polarization amplitude P (μC/cm²) and dielectric permittivity $\varepsilon_r$ (dimensionless), based on the extended Hamiltonian model described in Section 3.1. The contour plot illustrates the nonlinear dependence of Josephson coupling on the barrier's ferroelectric properties. The simulated range of 1−9 μC/cm² and $\varepsilon_r$=200-1000 covers the experimentally accessible regime for BNT thin films. Higher P and $\varepsilon_r$ values enhance the effective coupling, leading to increased $I_c$. This sensitivity analysis supports the fitting performed in the main text and confirms consistency between model parameters and physical properties of the BNT barrier under THz-field excitation*

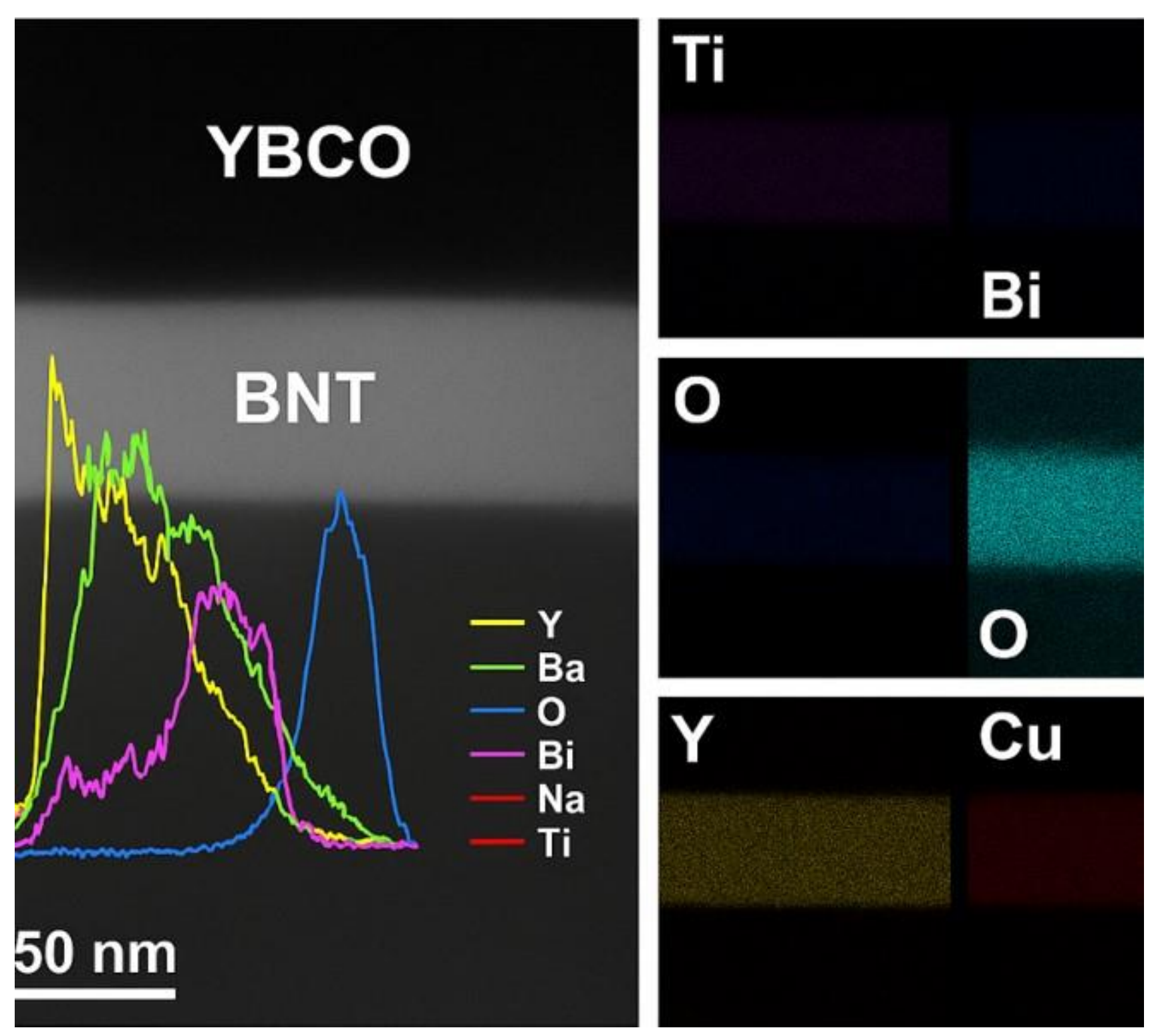

***Figure S3. High-resolution cross-sectional TEM and EDS elemental mapping of the YBCO|BNT|YBCO junction.***

*The TEM image confirms the structural continuity and uniformity of the trilayer stack. EDS elemental maps for Bi, Na, Ti (BNT layer) and Y, Ba, Cu (YBCO layers) demonstrate sharp compositional interfaces with no detectable interdiffusion, pinholes, or barrier discontinuities. These observations verify the compositional integrity and spatial uniformity of the 40 nm-thick BNT layer, excluding defect-mediated transport paths.*

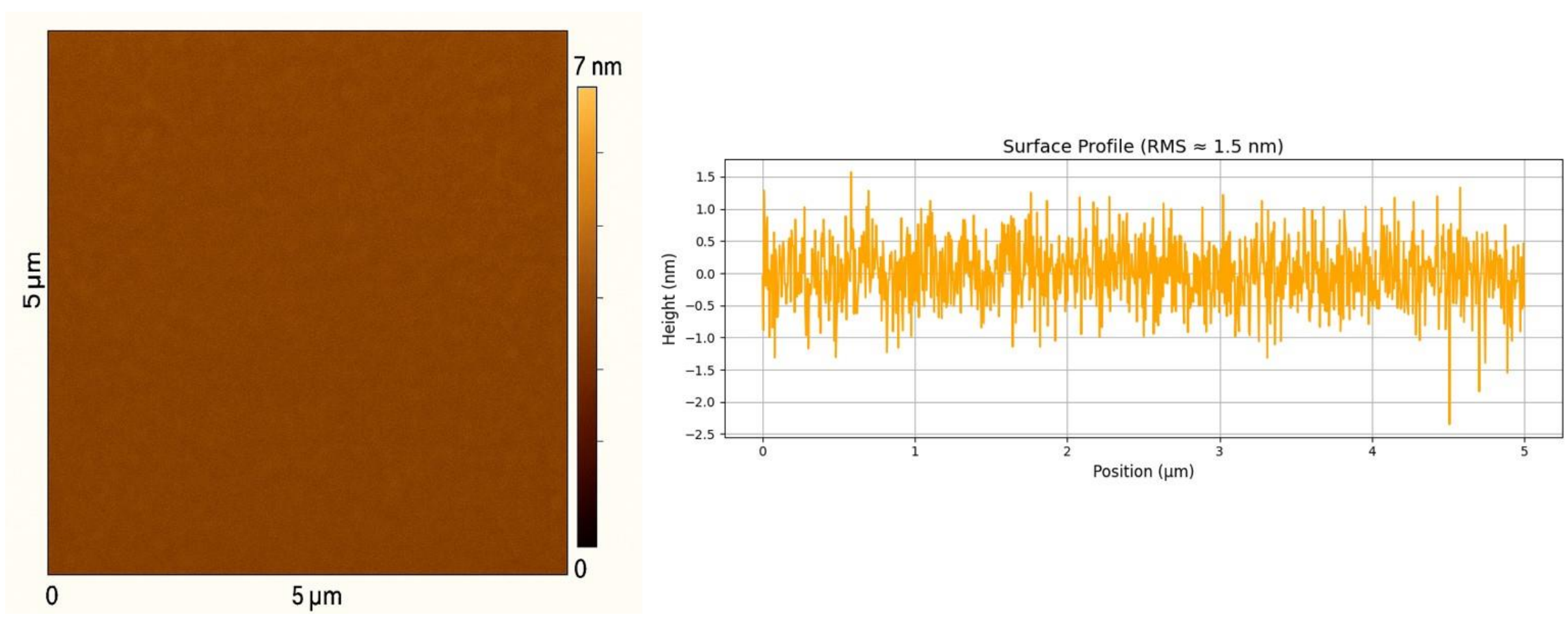

***Fig. S4. Atomic Force Microscopy (AFM) image of the BNT surface prior to top YBCO deposition.***

*Topographical scan over a 5 × 5 μm² area shows RMS surface roughness below 1.5 nm, indicating high smoothness and uniformity of the BNT barrier. Such low roughness supports coherent YBCO overlayer deposition and reduces the likelihood of local shorting or filamentary conduction through the barrier*

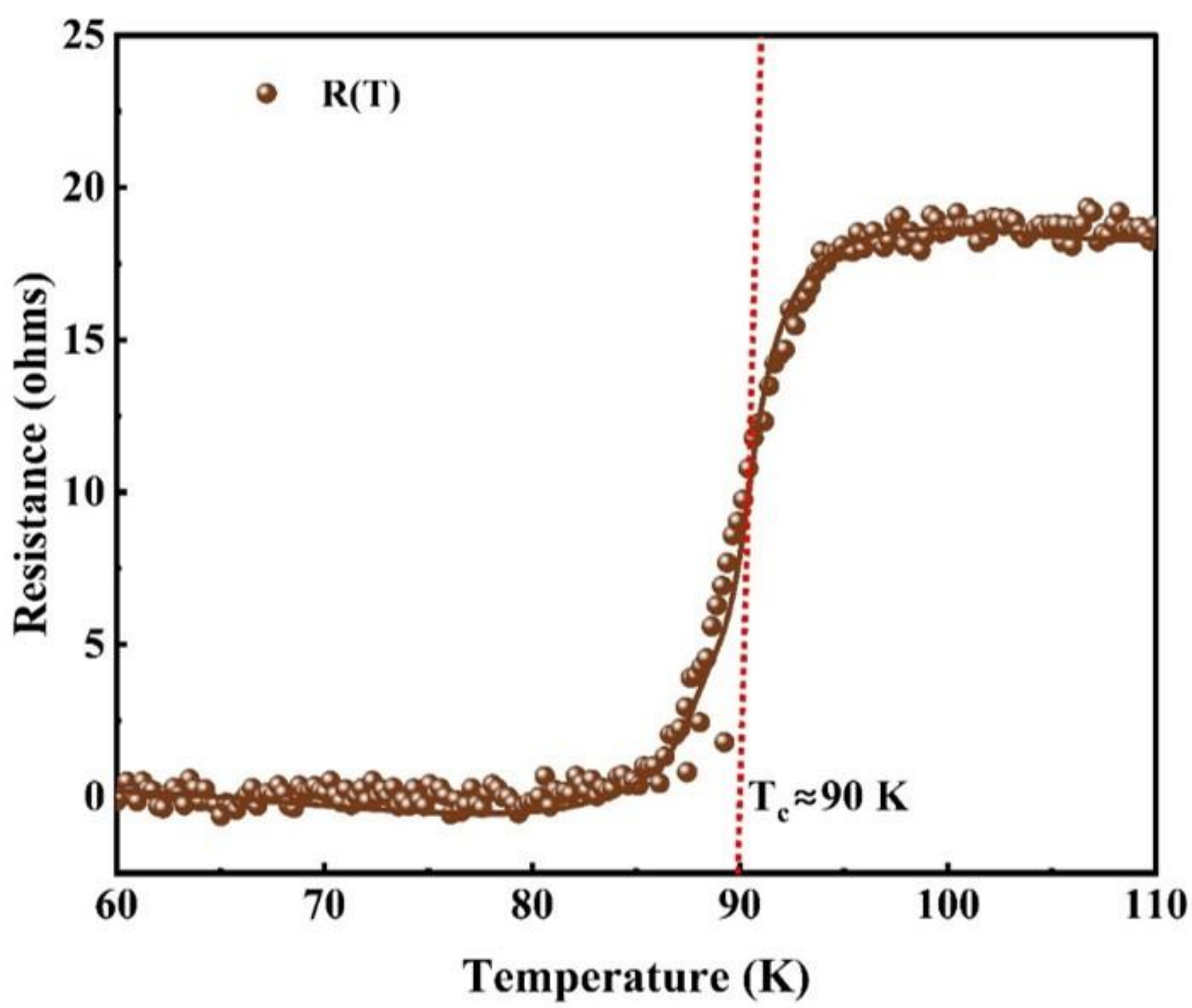

***Figure S5.*** *Temperature-dependent resistance R(T) of a standalone YBCO base electrode. The data show a sharp superconducting transition with the resistance dropping below the measurement resolution near the critical temperature,* $T_c \approx 90$ *(marked by the dashed red line). The well-defined transition confirms the high-quality superconducting phase of the YBCO layer.*

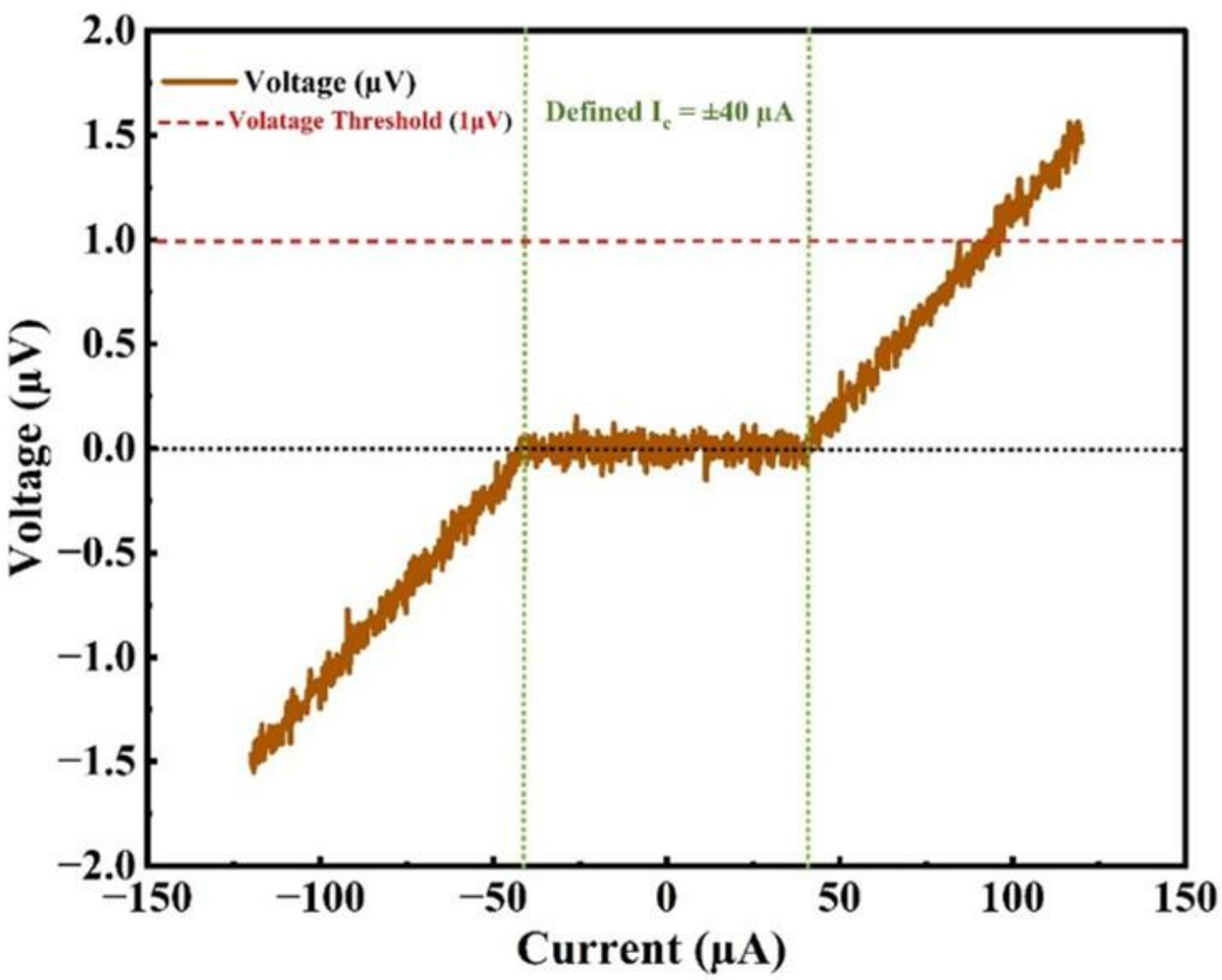

***Figure S6.*** *Current–voltage (I–V) characteristics of the YBCO|BNT|YBCO trilayer junction measured at 10 K. A well-defined zero-voltage plateau is observed for currents within approximately ±40 μA, consistent with coherent Cooper pair tunneling. The critical current ($I_c$) is determined using the ±1 μV threshold criterion (red dashed line), yielding $I_c$≈±40 μA, as indicated by the green dotted lines. The sharp transition from the zero-voltage state to the resistive branch is characteristic of Josephson junction behavior.*

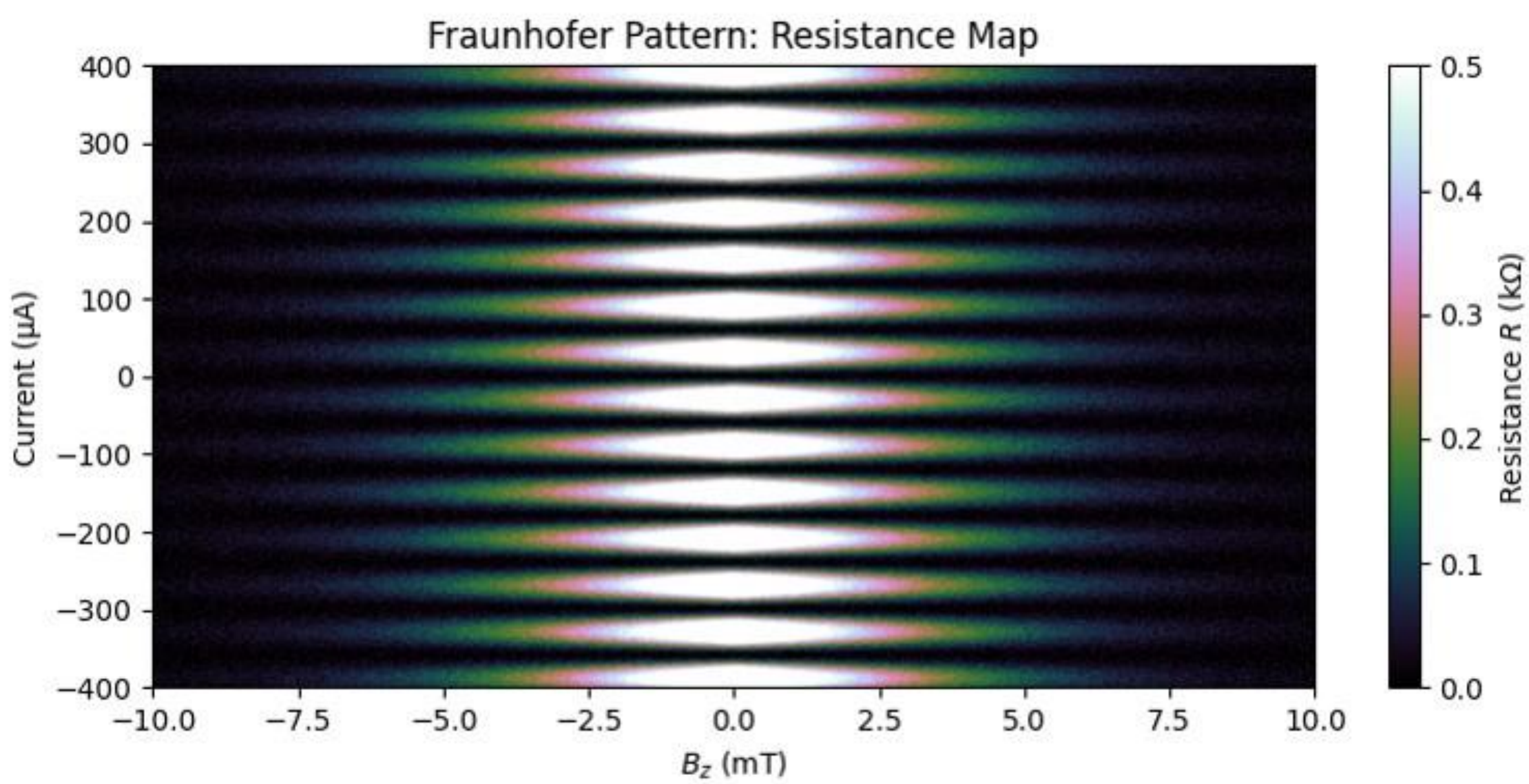

***Figure S7.*** *Two-dimensional resistance map of the YBCO|BNT|YBCO trilayer as a function of bias current and out-of-plane magnetic field (Bz). The alternating lobes of enhanced and suppressed resistance form a Fraunhofer-like interference pattern, with a central maximum at zero field and progressively weaker side lobes. This characteristic modulation confirms coherent Josephson coupling across the trilayer junction.*

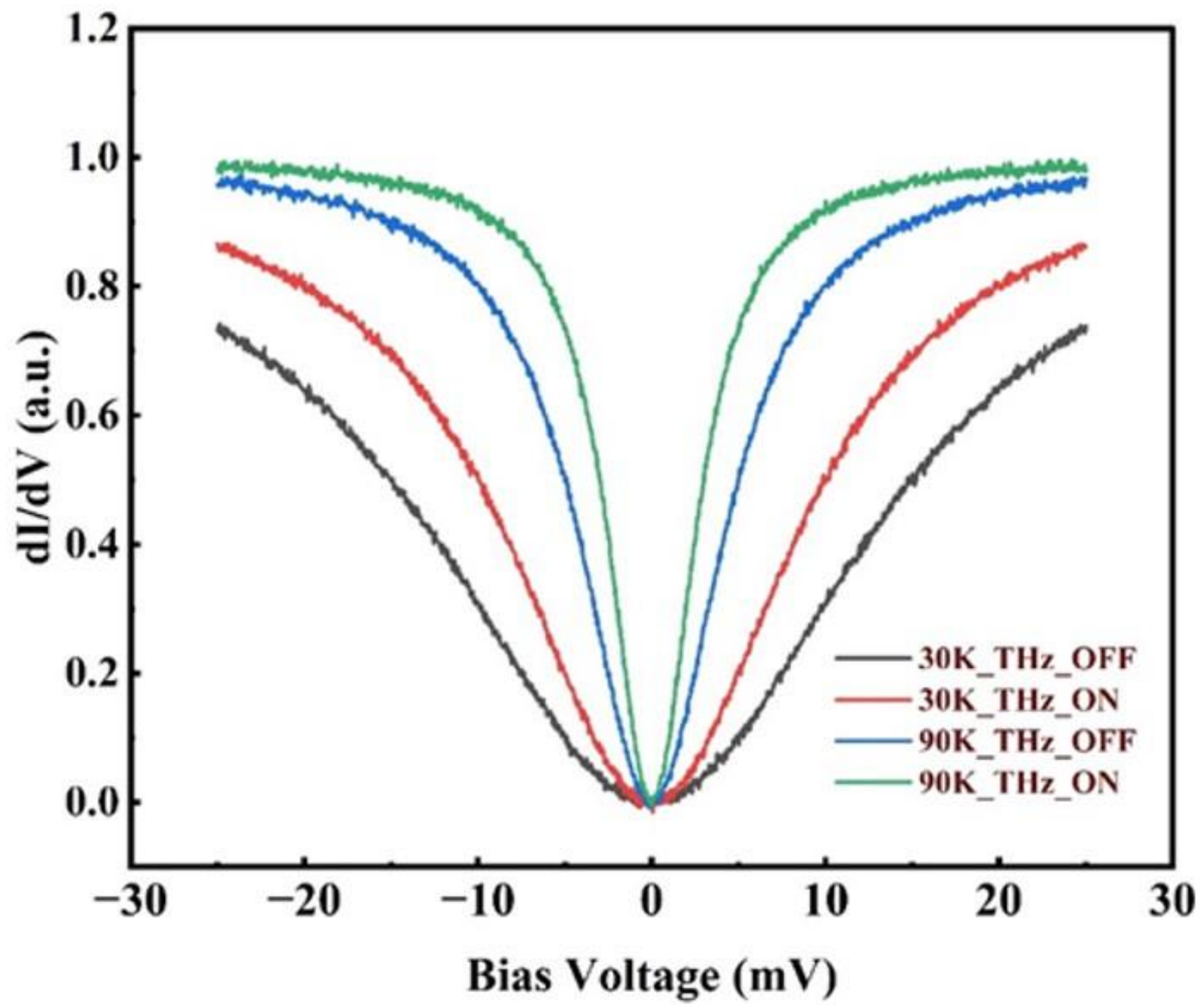

***Figure S8 | STM tunneling spectra under THz excitation.***

*Normalized dI/dV spectra recorded at 30 K and 90 K with and without a 1.5 THz field. At 30 K, the spectra under THz excitation show reversible suppression of coherence peaks and partial filling of the superconducting gap, consistent with polarization-mediated modulation of $\Delta$. No significant heating was observed, confirming the non-thermal nature of the effect.*

| Parameter | Symbol | Value / Range | Notes |
|---|---|---|---|
| Dielectric Permittivity | $\varepsilon_r$ | 200 – 1000 | Frequency- and temperature-dependent; consistent with BNT literature [1-3] |
| Polarization Amplitude | $P(\omega,T)$ | 1 – 9 μC/cm² | Varies with THz frequency and temperature |
| Josephson Coupling Constant | $J_0$ | Normalized to 1 (arb. units) | Sets baseline $I_c$ without THz field |
| THz Field Strength | $E_{THz}$ | 20 – 80 kV/cm | Controlled in experiment |
| Characteristic Frequency | $\omega_0$ | 0.35 – 0.45 THz | Center of resonant response |
| Damping Factor | $\gamma$ | 0.05 – 0.15 (arb. units) | Accounts for ferroelectric loss and damping |

***Table S1***. *Fitted Model Parameters for THz-Driven Josephson Junctions (Used in Eq. 5)*

| **Temperature (K)** | **σ(ω) ($10^4 \Omega^{-1} cm^{-1}$)** | **Δ(T) (meV)** |
|---|---|---|
| **30** | **1.21** | **1.77** |
| **40** | **0.95** | **1.52** |
| **50** | **0.76** | **1.25** |
| **60** | **0.66** | **1.02** |
| **80** | **0.54** | **0.58** |
| **90** | **0.42** | **0.00** |

***Table 2****: Calculated the superconducting energy gaps Δ(T) using Eq.13, at various temperatures*